\documentclass[useAMS,usenatbib]{mn2e}
\usepackage{natbib}
\usepackage{amsmath,amsfonts}
\usepackage{epsfig}

\def\reff@jnl#1{{\rm#1\/}}

\def\aj{\reff@jnl{AJ}}                  
\def\araa{\reff@jnl{ARA\&A}}            
\def\apj{\reff@jnl{ApJ}}                        
\def\apjl{\reff@jnl{ApJ}}               
\def\apjs{\reff@jnl{ApJS}}              
\def\ao{\reff@jnl{Appl.Optics}}         
\def\apss{\reff@jnl{Ap\&SS}}            
\def\aap{\reff@jnl{A\&A}}               
\def\aapr{\reff@jnl{A\&A~Rev.}}         
\def\aaps{\reff@jnl{A\&AS}}             
\def\azh{\reff@jnl{AZh}}                        
\def\baas{\reff@jnl{BAAS}}              
\def\jrasc{\reff@jnl{JRASC}}            
\def\memras{\reff@jnl{MmRAS}}           
\def\mnras{\reff@jnl{MNRAS}}            
\def\pra{\reff@jnl{Phys. Rev. A}}         
\def\prb{\reff@jnl{Phys. Rev. B}}         
\def\prc{\reff@jnl{Phys. Rev. C}}         
\def\prd{\reff@jnl{Phys. Rev. D}}         
\def\prl{\reff@jnl{Phys. Rev. Lett}}      
\def\pasp{\reff@jnl{PASP}}              
\def\pasj{\reff@jnl{PASJ}}              
\def\qjras{\reff@jnl{QJRAS}}            
\def\skytel{\reff@jnl{S\&T}}            
\def\solphys{\reff@jnl{Solar~Phys.}}    
\def\sovast{\reff@jnl{Soviet~Ast.}}     
\def\ssr{\reff@jnl{Space~Sci.Rev.}}     
\def\zap{\reff@jnl{ZAp}}                        
\def\nat{\reff@jnl{Nature}}             
\newcommand{\bvec}[1]{\mbox{\boldmath $#1$}}
\newcommand{\sbvec}[1]{\mbox{\boldmath $\scriptstyle #1$}}
\def\p#1by#2{{\partial{#1} \over \partial{#2}}}
\def\pp#1by#2#3{{\partial^2{#1} \over \partial{#2}\partial{#3}}}
\def\d#1by#2{{{\rm d}{#1} \over {\rm d}{#2}}}
\def\dd#1by#2#3{{{\rm d}^2{#1} \over {\rm d}{#2}{\rm d}{#3}}}

\title[CMB observations from the VSA \& CBI]
{CMB observations from the CBI and VSA: A comparison of coincident maps and parameter estimation methods}

\label{firstpage}
\author[Nutan Rajguru et al.] 
{Nutan Rajguru$^{1,\star}$,
 Steven T. Myers$^2$,  
 Richard A. Battye$^3$,
 J. Richard Bond$^4$, 
\newauthor 
 Kieran Cleary$^{3,\diamond}$, 
 Carlo R. Contaldi$^4$, 
 Rod D. Davies$^3$, 
 Richard J. Davis$^3$, 
\newauthor 
 Clive Dickinson$^{3,5}$, 
 Ricardo Genova-Santos$^6$, 
 Keith Grainge$^1$, 
 Yaser A. Hafez$^3$, 
\newauthor
 Michael P. Hobson$^1$, 
 Michael E.  Jones$^{1,\ddagger}$, 
 R\"udiger Kneissl$^{1,\circ}$, 
 Katy Lancaster$^{1,+}$, 
\newauthor 
 Anthony Lasenby$^1$, 
 Brian S. Mason$^7$,
 Timothy J. Pearson$^5$, 
 Guy G. Pooley$^1$, 
\newauthor  
 Anthony C. S. Readhead$^5$,
 Rafael Rebolo$^{6}$,
 Graca Rocha$^1$,
 Jos\'e Alberto Rubi\~no-Martin$^{6}$, 
\newauthor 
 Richard D. E. Saunders$^1$, 
 Richard S. Savage$^{1,\dagger}$, 
 Anna Scaife$^1$,
 Paul F. Scott$^1$, 
\newauthor
 Jonathan. L. Sievers$^4$,
 An\v{z}e Slosar$^{1,\times}$, 
 Angela C. Taylor$^{1,\ddagger}$,
 David Titterington$^1$,  
 \newauthor 
 Elizabeth Waldram$^1$,
 Robert A. Watson$^3$, 
 Althea Wilkinson$^3$. \vspace{0.03in}\\
$^1$  Astrophysics Group, Cavendish Laboratory, Madingley Road, Cambridge CB3 0HE, UK.\\
$^2$ National Radio Astronomy Observatory, Socorro, NM 87801, USA.\\
$^3$ Jodrell Bank Observatory, Macclesfield, Cheshire SK11 9DL, UK. \\
$^4$ Canadian Institute for Theoretical Astrophysics, University of Toronto, 60 St. George Street, Toronto, Ontario, M5S 3H8, Canada. \\
$^5$ California Institute of Technology, Mail Code 105-24, Pasadena, CA 91125, USA.\\
$^6$ Instituto de Astrof{\'i}sica de Canarias, 38200 La Laguna, Tenerife, Spain.\\ 
$^7$ National Radio Astronomy Observatory, PO Box 2, Green Bank, WV 24944, USA.\\
$^{\diamond}$ Present address: Jet Propulsion Laboratory, 4800 Oak Grove Drive, Pasadena, CA 91109, USA.\\
$^{\ddagger}$ Present address: Department of Physics, The Denys Wilkinson Building, Keble Road, Oxford, OX1 3RH, UK.\\
$^{\circ}$ Present address: Department of Physics, University of California, Berkeley, CA 94720-7300, USA. \\
$^{+}$ Present address: HH Wills Physics Laboratory, Tyndall Avenue, Bristol BS8 1TL, UK. \\
$^{\dagger}$ Present address: Astronomy Centre, University of Sussex, Brighton, BN1 9QH, UK.\\
$^{\times}$ Present address: Faculty of Mathematics and Physics, University of Ljubljana, 1000 Ljubljana, Slovenia.\\
$^{\star}$ E-mail: nr243@mrao.cam.ac.uk \\
}

\date{Accepted ---; received ---; in original form \today}
\pagerange{\pageref{firstpage}--\pageref{lastpage}}
\pubyear{2005}

\begin{document}
\maketitle

\begin{abstract}
We present coincident observations of the Cosmic Microwave Background (CMB) from the Very Small Array (VSA) and Cosmic Background Imager (CBI) telescopes. The consistency of the full datasets is tested in the map plane and the Fourier plane, prior to the usual compression of CMB data into flat bandpowers. Of the three mosaics observed by each group, two are found to be in excellent agreement. In the third mosaic, there is a 2$\sigma$ discrepancy between the correlation of the data and the level expected from Monte Carlo simulations. This is shown to be consistent with increased phase calibration errors on VSA data during summer observations. We also consider the parameter estimation method of each group. The key difference is the use of the variance window function in place of the bandpower window function, an approximation used by the VSA group. A re-evaluation of the VSA parameter estimates, using bandpower windows, shows that the two methods yield consistent results.
\end{abstract}


\begin{figure*}
\includegraphics[width=0.347\textwidth]{./figures/fig1_vsauv.ps}\hspace{0.14in}
\includegraphics[width=0.365\textwidth]{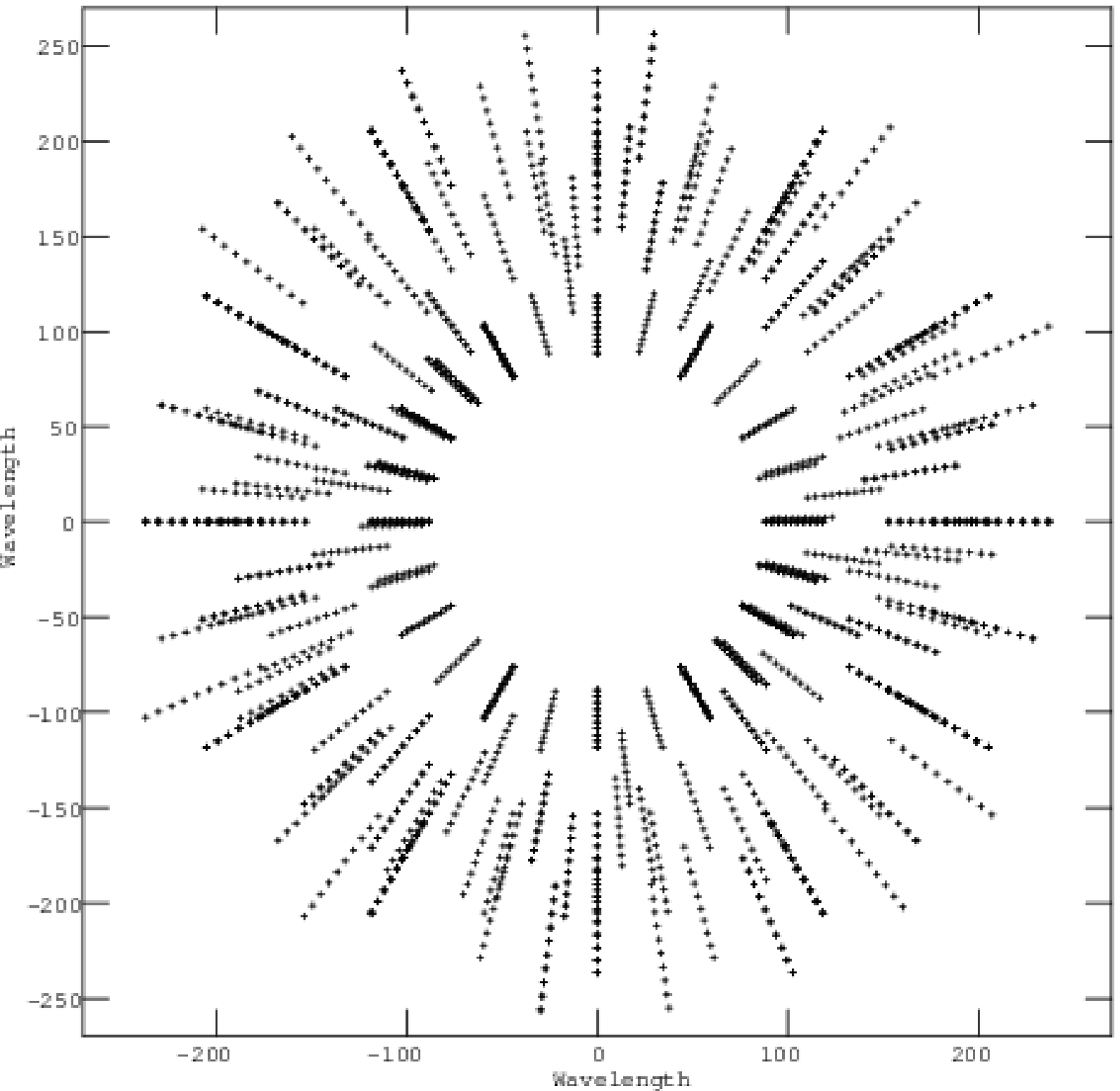}
\includegraphics[width=0.38\textwidth]{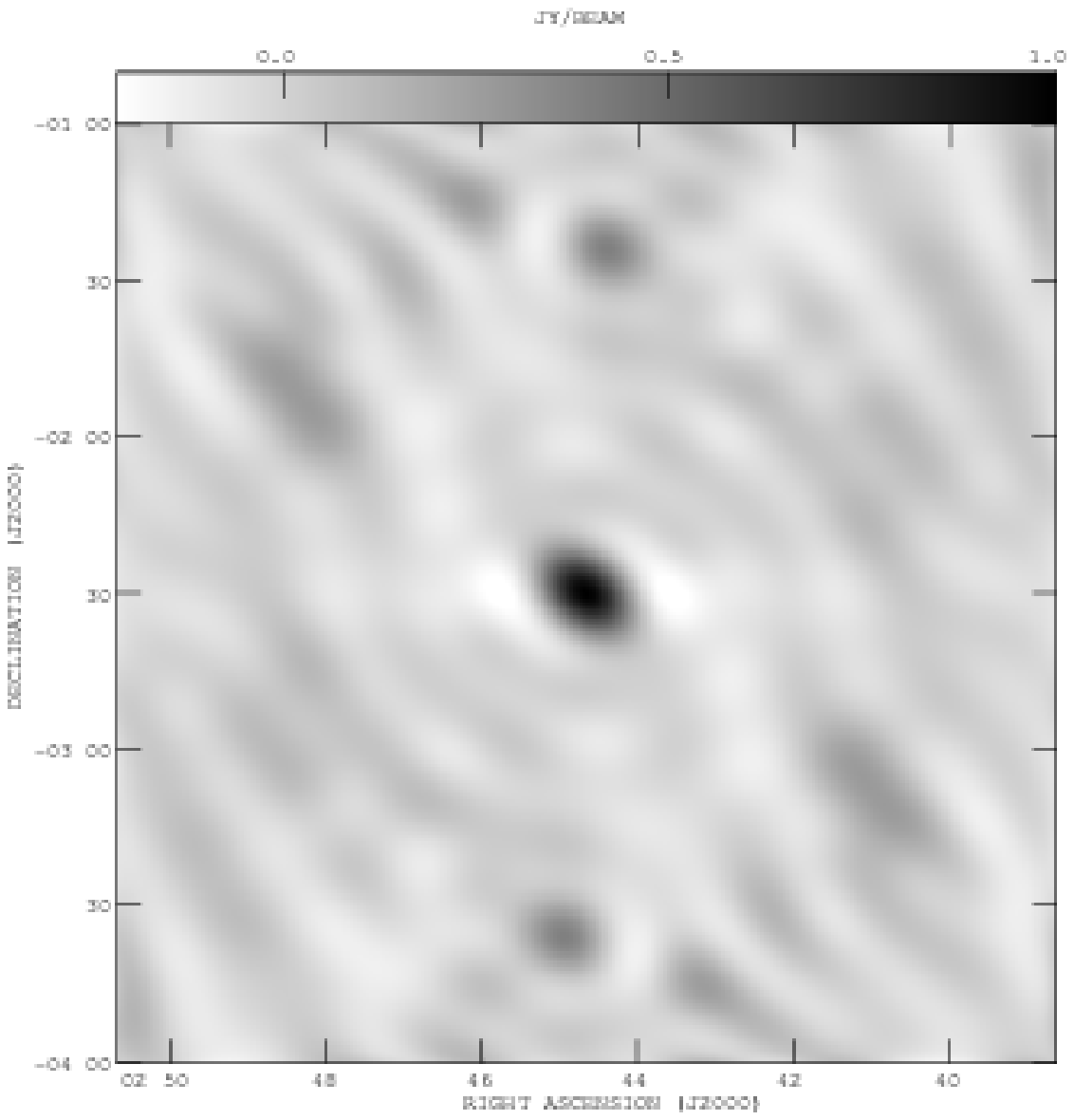}\hspace{0.27in}
\hspace{0.12in}\includegraphics[width=0.32\textwidth]{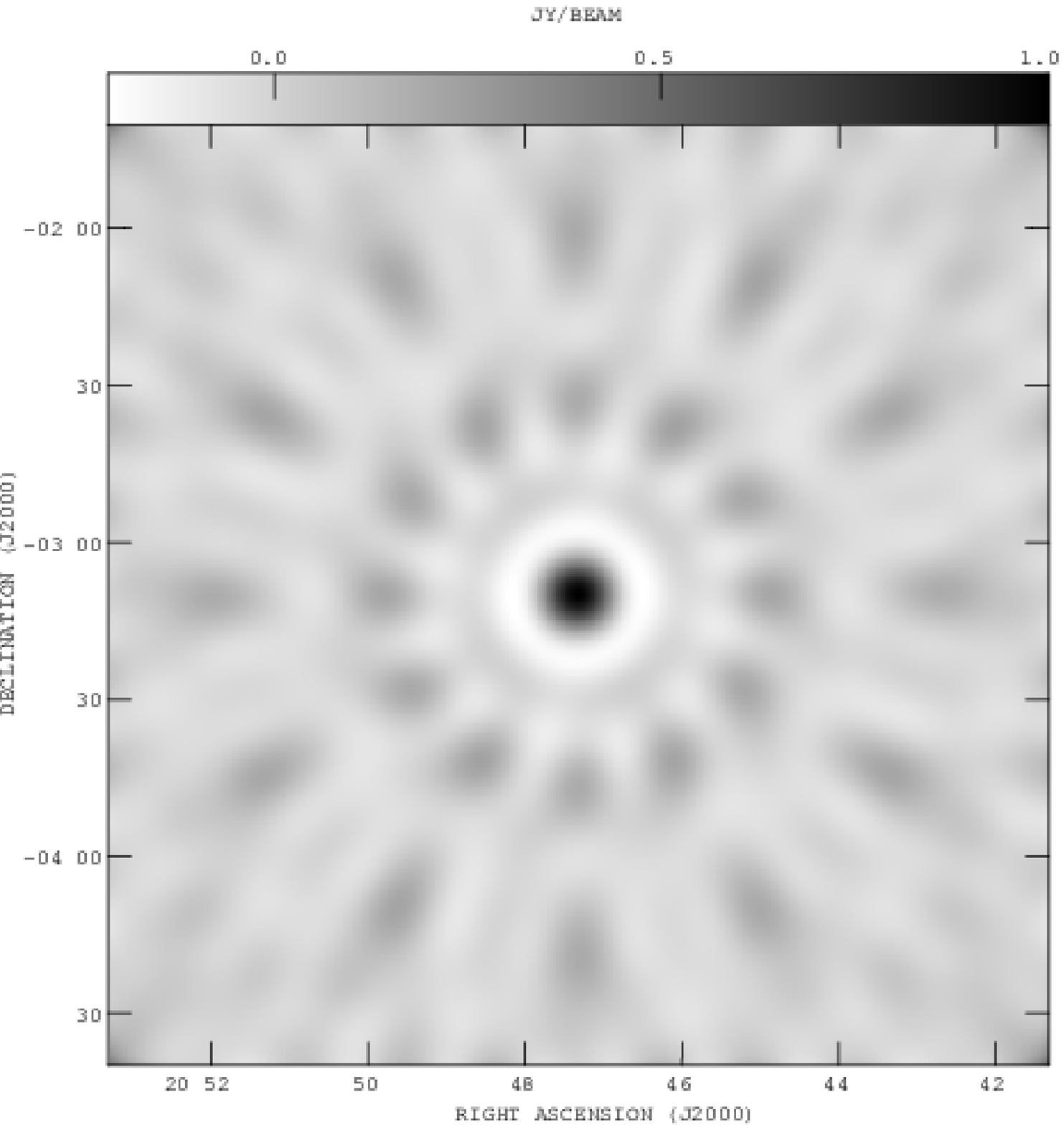}
\caption{Typical uv-coverage. {\it Top left}: VSA. Each point represents the full 
1.5 GHz bandwidth. {\it Top right}: CBI. Each point represents a single 1 GHz 
channel. {\it Bottom}: Synthesised beams for typical uv-coverage.} \label{fig:uv}
\end{figure*}

\section{Introduction}

Measurements of cosmic microwave background (CMB) anisotropies have proved invaluable in establishing the current $\Lambda$CDM model and are widely used in cosmological parameter estimation \citep{2003ApJS..148..175Sshort,2003ApJ...599..773Gshort,2004MNRAS.353..747Rshort, 2004ApJ...609..498Rshort}. An important check on the accuracy of CMB measurements is to compare the data obtained from instruments which are subject to different systematic effects. The CBI and VSA instruments have significant design differences and the comparison of our data is a check that known systematic effects have been accurately corrected for, and that neither dataset is seriously contaminated by unrecognised systematic errors. In the CMB community, the focus has been on the comparison of power spectra. However, maps of CMB anisotropies also contain important information. In particular, they are used in tests of Gaussianity \citep[see for example,][]{2003ApJS..148..119Kshort,2003NewAR..47..821A,2004MNRAS.349..973Sshort}, a key assumption of CMB analysis. It is therefore an important exercise to check the correlation between maps of CMB anisotropies. To this end the VSA has undertaken a programme of observing fields previously imaged by the CBI. In this paper we present the results of these observations and assess their consistency with the CBI data.

Key scientific results from the measurement of CMB anisotropies are the cosmological parameters. An essential ingredient in converting the flat bandpower estimates into cosmological parameters are the window functions which allow a theoretical power spectrum to predict a flat bandpower. Both the CBI and VSA groups use a maximum likelihood method of estimating the bandpowers, with some differences in the implementation, but there are important differences in the type of window functions used for parameter estimation. The CBI group computes the bandpower window function which fully takes into account the anti-correlations between neighbouring bins \citep{2003ApJ...591..575Mshort}. The VSA group use variance windows as an approximation to the bandpower windows when computing parameter estimates \citep{2003MNRAS.341.1084Rshort,2003MNRAS.341L..29Sshort,2004MNRAS.353..747Rshort}, although anti-correlations are not accounted for. We assess the bias resulting from this approximation by  re-evaluating VSA parameter estimates using bandpower window functions.

In section 2 we summarise the key differences between the VSA and CBI instruments and the implications for observing strategies. In section 3 we describe how the differenced maps are produced and present the results of the comparison. Section 4 contains our assessment of the impact of  window functions on the VSA parameter estimation. Finally, in section 5 we present our conclusions.

\begin{table*}
\caption{A summary of the specifications of the CBI and the VSA extended array. 
\label{tab:spec}}
\begin{tabular}{lcc}
\hline \hline
& VSA & CBI\\
\hline
Observing Frequency             &33\,GHz                  &31.5\,GHz\\
Bandwidth                       &1.5\,GHz                 &10\,GHz\\
Number of Channels              &1                       &10\\
Number of Antennas              &14                      &13\\
Number of Baselines             &91                      &78\\
Range of baseline lengths	&$0.6$\,m$- 2.5$\,m        &$1.0$\,m$- 5.5$\,m\\
$\ell$ range			&$\approx 300-1500$         &$\approx 300-3500$\\
Primary Beam (FWHM)             &$2.11^{\circ}$  &45.2 arcmin $\times$ 31\,GHz/$\nu$\\              
System temperature      	&$\approx 35$\,K             &$\approx 30$\,K\\
Mirror diameters		&0.32\,m                 &0.90\,m\\
Synthesised beam (FWHM)		&$\approx 11$ arcmin        &$\approx5$ arcmin\\ 
Flux sensitivity                &50\,Jy\,s$^{-1/2}$         &1.5\,Jy\,s$^{-1/2}$\\
\hline \hline
\end{tabular}
\end{table*}

\section{The Telescopes}\label{telescopes}

The CBI is an interferometric telescope located at an altitude of 5000\,m in the Atacama Desert in northern Chile. The instrument operates in ten 1-GHz frequency bands over 26$-36$\,GHz. The antennas have low-noise high electron mobility transistor (HEMT) amplifiers and  typical system temperatures including the CMB, ground and atmosphere are $\approx$30\,K. The 13 Cassegrain antennas, each 0.9\,m in diameter are co-mounted on a 6\,m tracking platform \citep{2002PASP..114...83Pshort}. 

The VSA is sited at the Teide Observatory, in Tenerife, at an altitude of 2400\,m. The VSA operates in a single 1.5-GHz channel at a central frequency of 33\,GHz. The 14 antennas also have HEMT amplifiers and the typical system temperature is $\approx$35\,K. In the extended array configuration, the VSA uses mirrors of diameter 0.322\,m. The VSA horn-reflector antennas are mounted on a tilt table hinged along east-west, but each antenna individually tracks the observed field by rotating its horn axis perpendicularly to the table hinge, and wavefront coherence is maintained with an electronic path compensator system \citep{2003MNRAS.341.1057Wshort}. This is a key design difference to the CBI. 

The individual tracking of the VSA antennas allows for the filtering of contaminating signals. These may be celestial sources such as the Sun and Moon, or ground-spill and other ground based spurious signals \citep{2003MNRAS.341.1057Wshort}. The fringe rates of contaminating signals differ sufficiently from those of the target to allow effective filtering whilst retaining most of the data. For example, filtering the Sun at a distance of 20$^{\circ}$ from the target removes approximately 25\% of the data. The VSA also uses a ground-shield to minimise ground-spill. The consequences of these design differences are two-fold. Firstly, the VSA is able to observe 24 hours a day and its extended array (as used here) can filter out emission from the Sun and Moon when they are as close as $9^{\circ}$. The CBI is limited to observing at night and to fields which are more than $60^{\circ}$ away from the Moon \citep{2002PASP..114...83Pshort}. Secondly, the VSA is unaffected by ground-spill contamination for fields within 35$^{\circ}$ of the zenith and so is able to make direct images of the sky. The raw CBI data are contaminated by ground-spill and this is eliminated by means of a differencing scheme. The method of differencing the CBI data used in this analysis works as follows. A lead field is observed followed by a trail field at the same declination but separated by 8 minutes in RA. The trail field visibilities are then subtracted from those of the lead field. This has the effect of removing the contaminating signal which is constant on an 8 minute timescale, whilst preserving the statistical distribution of the sky Fourier modes \citep{2002PASP..114...83Pshort}. Both sky maps and the power spectrum are estimated from the differenced data.

The CBI mounting platform allows the orientation of the baselines to be changed by rotating the platform about the optical axis. The rotation of the tracking table together with the broad bandwidth ensures a well-filled aperture plane and a circularly symmetric synthesised beam. Figure \ref{fig:uv} shows the typical uv-coverage and synthesised beams of each telescope for the observations presented in this paper, in the range of angular scales common to both experiments. The uv-coverage of the VSA is less complete and the beam is less circularly symmetric than that of the CBI, for these fields. This is partly due to the low declination of the observations ($-3.5^{\circ}$) which is close to the lowest elevation that the VSA is able to observe. The telescope is located at a latitude of $28^{\circ}$ and at the higher declinations of the main primordial fields, uv-coverage is significantly better \citep{2003MNRAS.341.1066Tshort,2004MNRAS.353..732Dshort}. Unlike the CBI, the VSA mounting table does not allow for rotation about the optical axis. Together with the lower bandwidth this limits the coverage of the aperture plane, although minimises the number of redundant baselines at a given frequency. A summary of the specification of the two telescopes is shown in Table \ref{tab:spec}.

\section{Observations}

The CBI data used in this comparison are the mosaics 02H, 14H and 20H and first reported in \cite{2003ApJ...591..556Pshort}. The CBI data also include two deep field observations which fall within the 14H and 20H mosaics and are described in \cite{2003ApJ...591..540Mshort}. Both CBI mosaiced and deep field data were collected during the period January - December 2000. Each mosaic consists of 42 differenced fields. 

The VSA observations were carried out at a later epoch. The data were made between May 2002 and May 2004 with the telescope in the extended array configuration. The larger primary beam size of the extended array mirrors allows each mosaic to be covered in three pointings. Table \ref{tab:fields} shows the coordinates and effective integration times for the VSA observations. The VSA data reduction and calibration procedure are described in \cite{2004MNRAS.353..732Dshort} and references therein.

\begin{table}
\begin{center}
\caption{Celestial coordinates for the VSA observations of the CBI 02H, 14H and 20H mosaics. The effective integration time is calculated after flagging and filtering of the data. 
\label{tab:fields}}
\begin{tabular}{lccc}
\hline \hline
\hspace{0.13in} Field & RA (J2000) & DEC (J2000) & $\rm t_{int}$ (hrs)\\
\hline
VSA-02H-a & 02 44 24.00 & -03 30 00.0 & 169.9\\
VSA-02H-b & 02 50 00.00 & -03 30 00.0 & 86.2\\
VSA-02H-c & 02 55 36.00 & -03 30 00.0 &  49.7\\
VSA-14H-a & 14 44 24.00 & -03 30 00.0 & 37.1\\
VSA-14H-b & 14 50 00.00 & -03 30 00.0 & 73.3\\
VSA-14H-c & 14 55 36.00 & -03 30 00.0 & 44.9\\
VSA-20H-a & 20 44 24.00 & -03 30 00.0 & 154.2\\
VSA-20H-b & 20 50 00.00 & -03 30 00.0 & 94.2\\
VSA-20H-c & 20 55 36.00 & -03 30 00.0 & 74.4\\
\hline \hline
\end{tabular}
\end{center}
\end{table}

\subsection{Foreground contamination}

At frequencies of 26-36\,GHz, the dominant cosmological contamination to CMB observations comes from galactic foregrounds and extragalactic radio sources. The diffuse galactic foregrounds include both bremsstrahlung and synchrotron emission. However, this emission is concentrated in the galactic plane and contamination may be avoided by observing at high galactic latitudes. The observations presented here are at galactic latitudes $>$ 20$^{\circ}$. In addition, the data are insensitive to the large angular scales where galactic contamination is significant. Therefore, the main contaminant is likely to be extragalactic point sources. 

The standard VSA source-subtraction strategy involves an initial survey with the Ryle Telescope (RT) at 15\,GHz \citep{2003MNRAS.342..915W}. Sources identified by the RT are then monitored with the source subtractor at 33\,GHz. The observations are carried out simultaneously with the CMB field observations to take account of the variability of the sources. A statistical correction is also applied to the power spectrum to remove the small effect of the remaining, fainter sources. As the RT is located in Cambridge at a latitude of +52$^{\circ}$, we were unable to survey fields at the low declinations of the CBI observations. For this reason, a more limited level of source subtraction was implemented. This involved the subtraction from the data of both groups, the fluxes obtained by the CBI group from observations at 31\,GHz with the 40-metre telescope at the Owens Valley Radio Observatory (OVRO). These observations were carried out, simultaneously, in so far as was possible, with the CBI observations. As noted above, the VSA observations were carried out a later epoch and no further source observations were carried out to account for the variability of sources.

The typical sensitivity achieved in the OVRO data was 2\,mJy (rms) which allows for a completeness estimate of 90\% at S$_{31} >$ 16\,mJy \citep{2003ApJ...591..540Mshort}. To achieve high $\ell$ measurements of the power spectrum, a deeper level of discrete source subtraction is required. To achieve this the usual approach of the CBI group is to employ the strategy of constraint matrices to `project out' sources at known positions but with unknown fluxes \citep{1998PhRvD..57.2117B}. All sources $>\,$3.4\,mJy in the 1.4\,GHz NVSS catalogue are projected out of the CBI data. A statistical correction based on the CBI source count is then applied for sources $<$\,3.4\,mJy at 1.4\,GHz \citep{2003ApJ...591..540Mshort}. For this investigation, the OVRO fluxes have been subtracted from the CBI data but the constraint matrix strategy has not been employed. This is due to the limited resolution of the data included in the comparison and to avoid removing a significant fraction of the data, which is unnecessary for this analysis.

The different methods usually employed to remove the effects of contaminating sources are a key distinction in the analysis of the two groups. Although we are unable to make a direct comparison of these strategies, due to the positions of the fields, the $33\,$GHz source counts estimated from the main VSA source monitoring programme have been used to evaluate the CBI source subtraction strategy. Based on VSA source counts \cite{cleary2004short} find the residual correction due to sources below the detection threshold of 3.4\,mJy at 1.4\,GHz to be $0.03\,\rm Jy^2\,\rm sr^{-1}$ which is consistent with the CBI group estimate of $0.08\,\pm\,0.04\,\rm Jy^2\,\rm sr^{-1}$.

\subsection{Maps}

\begin{figure*}
\includegraphics[width=0.41\textwidth,clip]{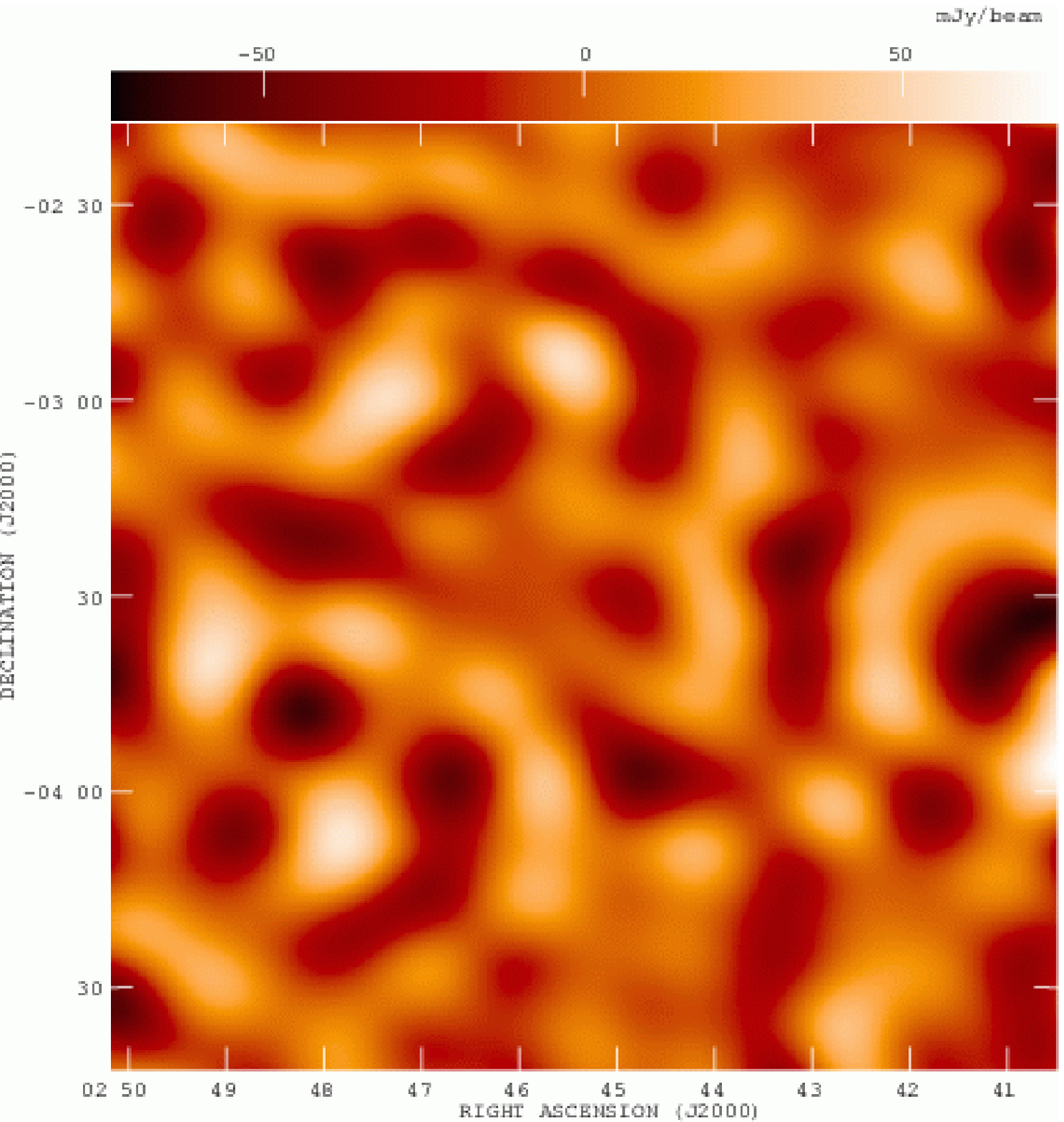}
\includegraphics[width=0.41\textwidth]{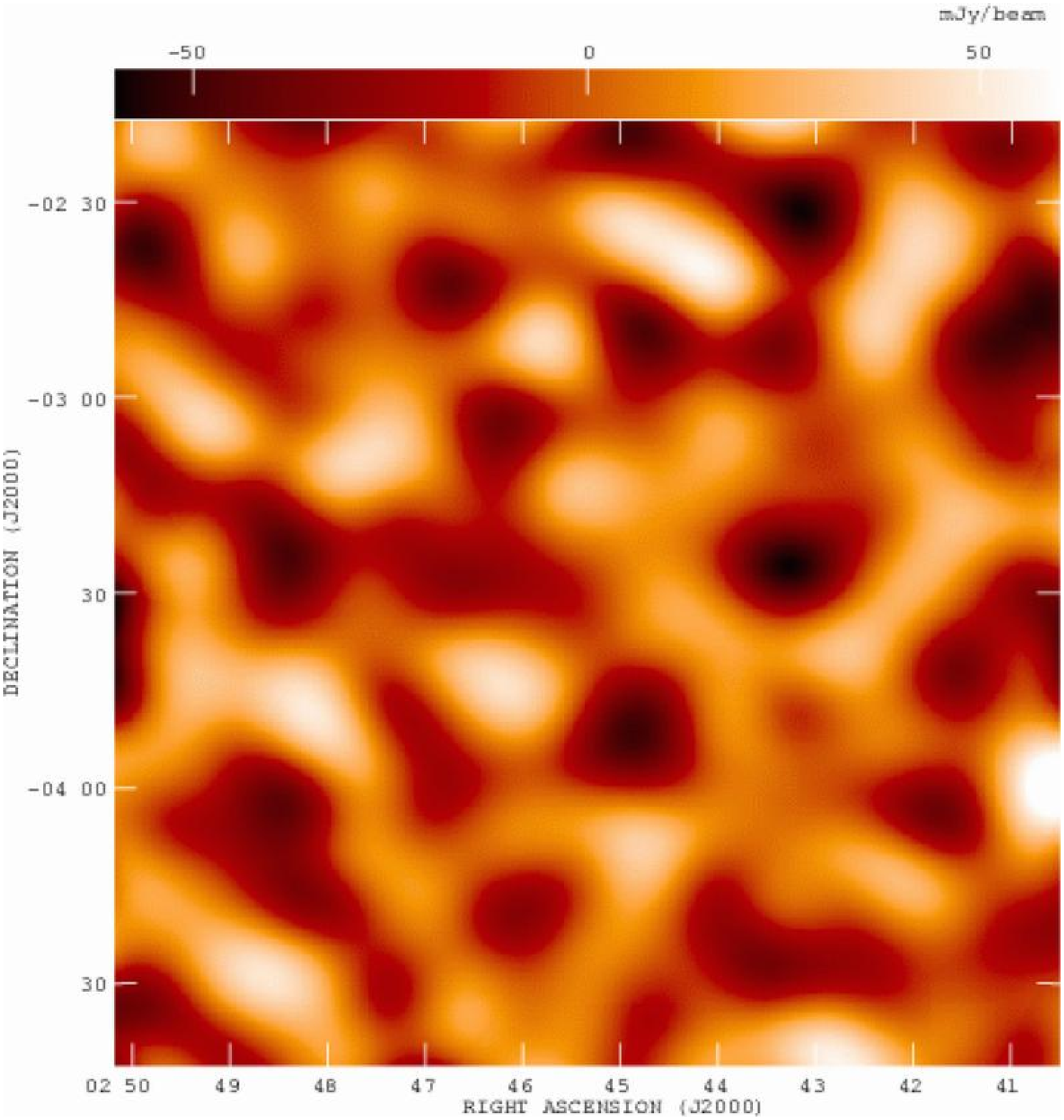}
\includegraphics[width=0.41\textwidth]{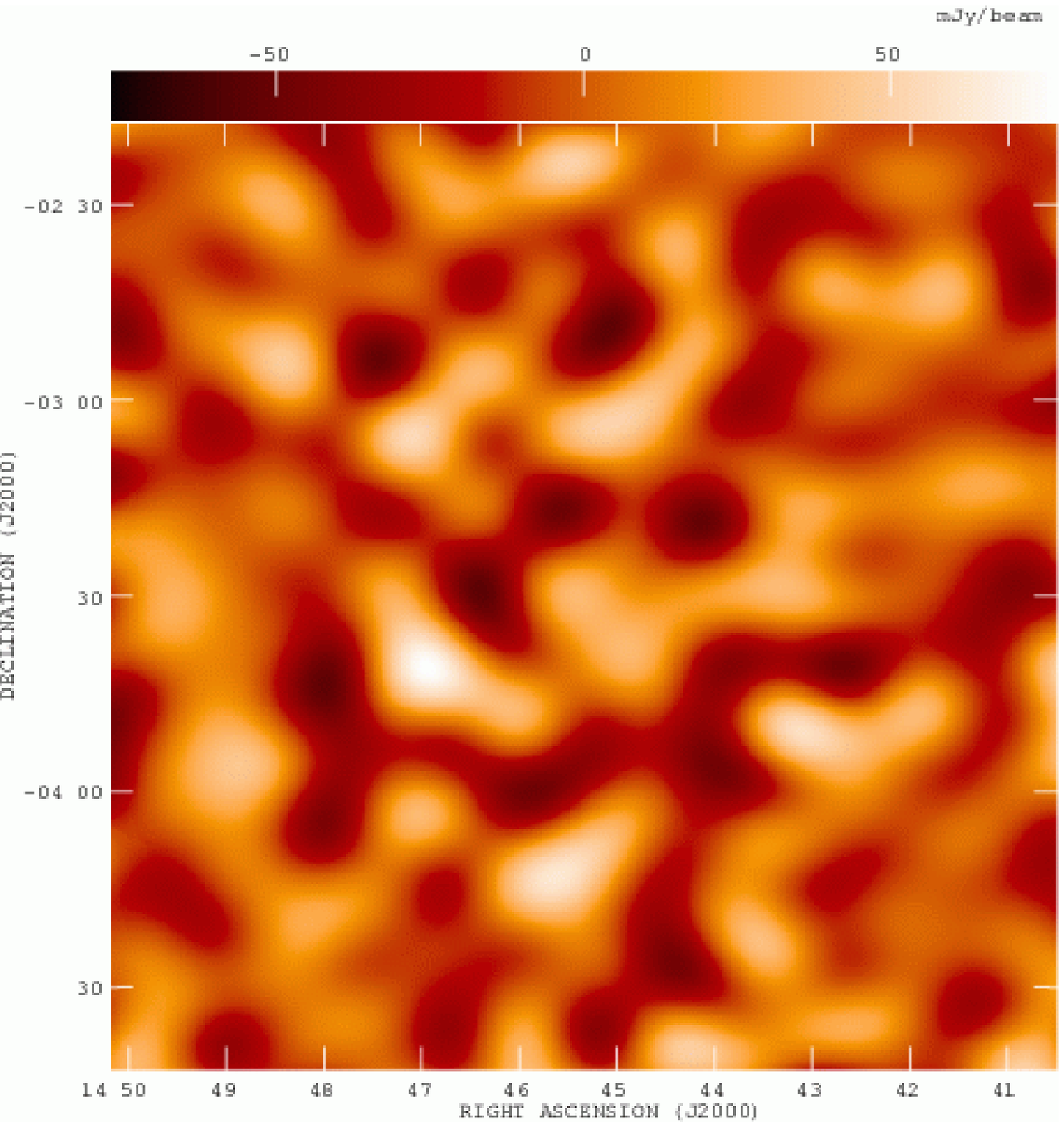}
\includegraphics[width=0.41\textwidth]{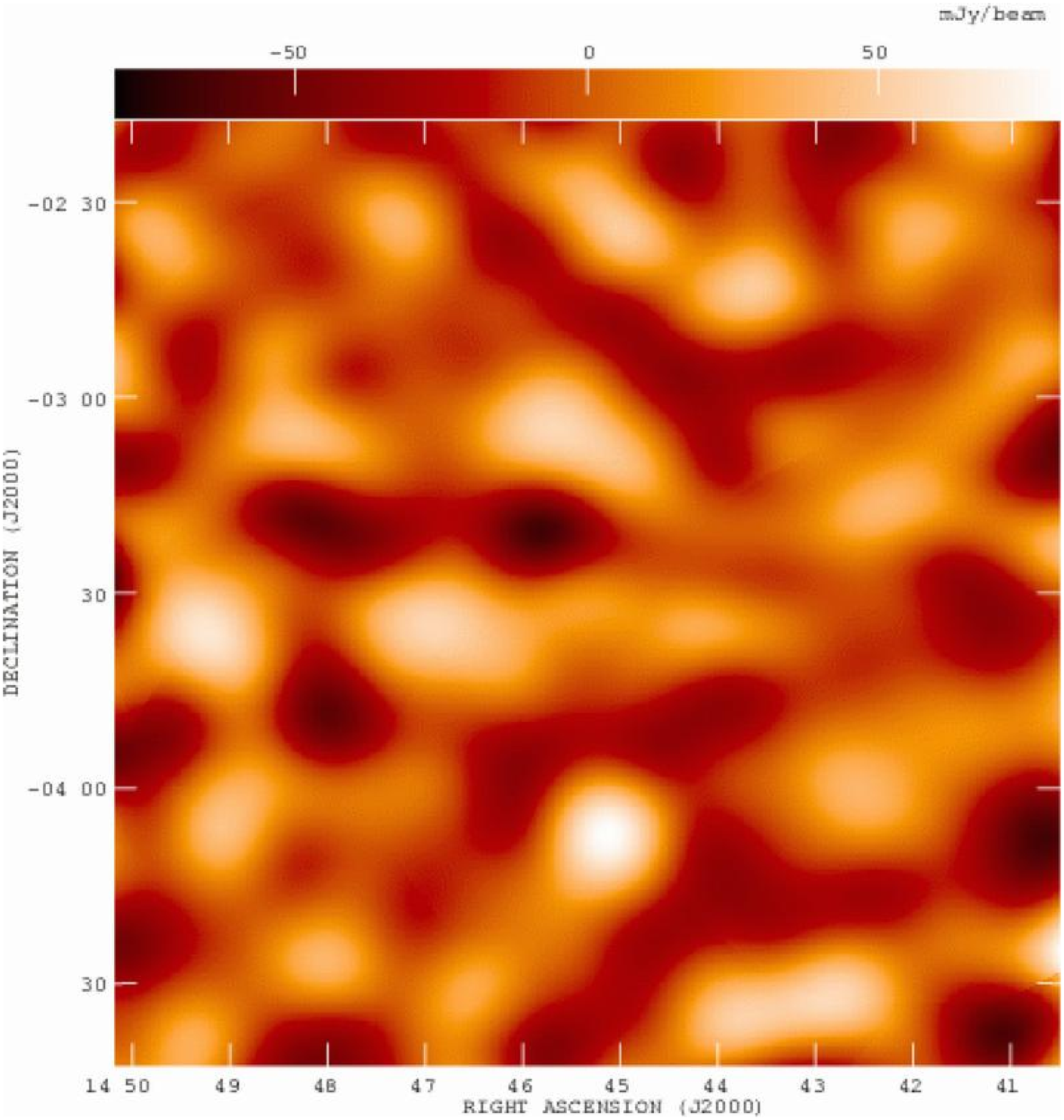}
\includegraphics[width=0.41\textwidth]{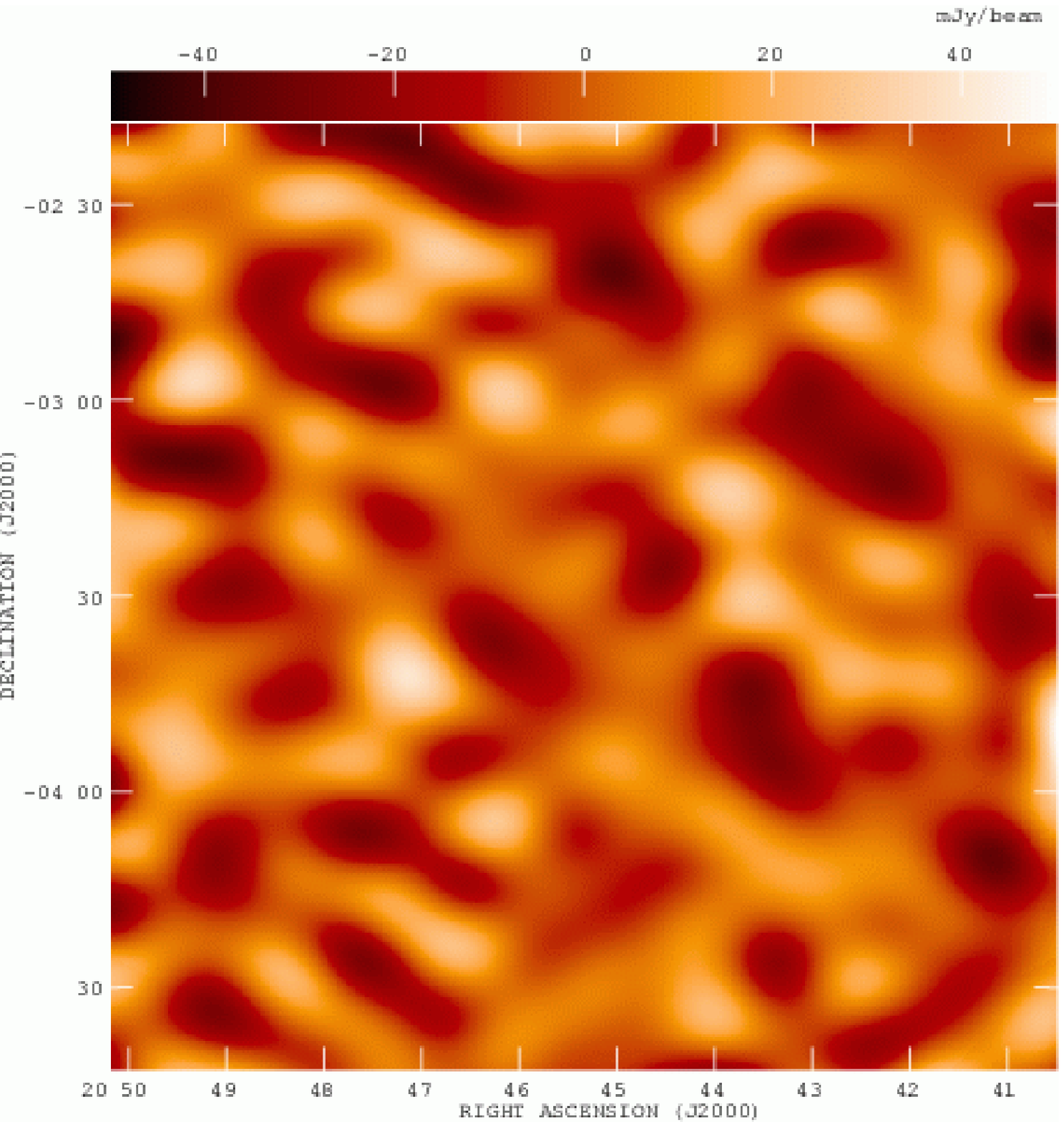}
\includegraphics[width=0.41\textwidth]{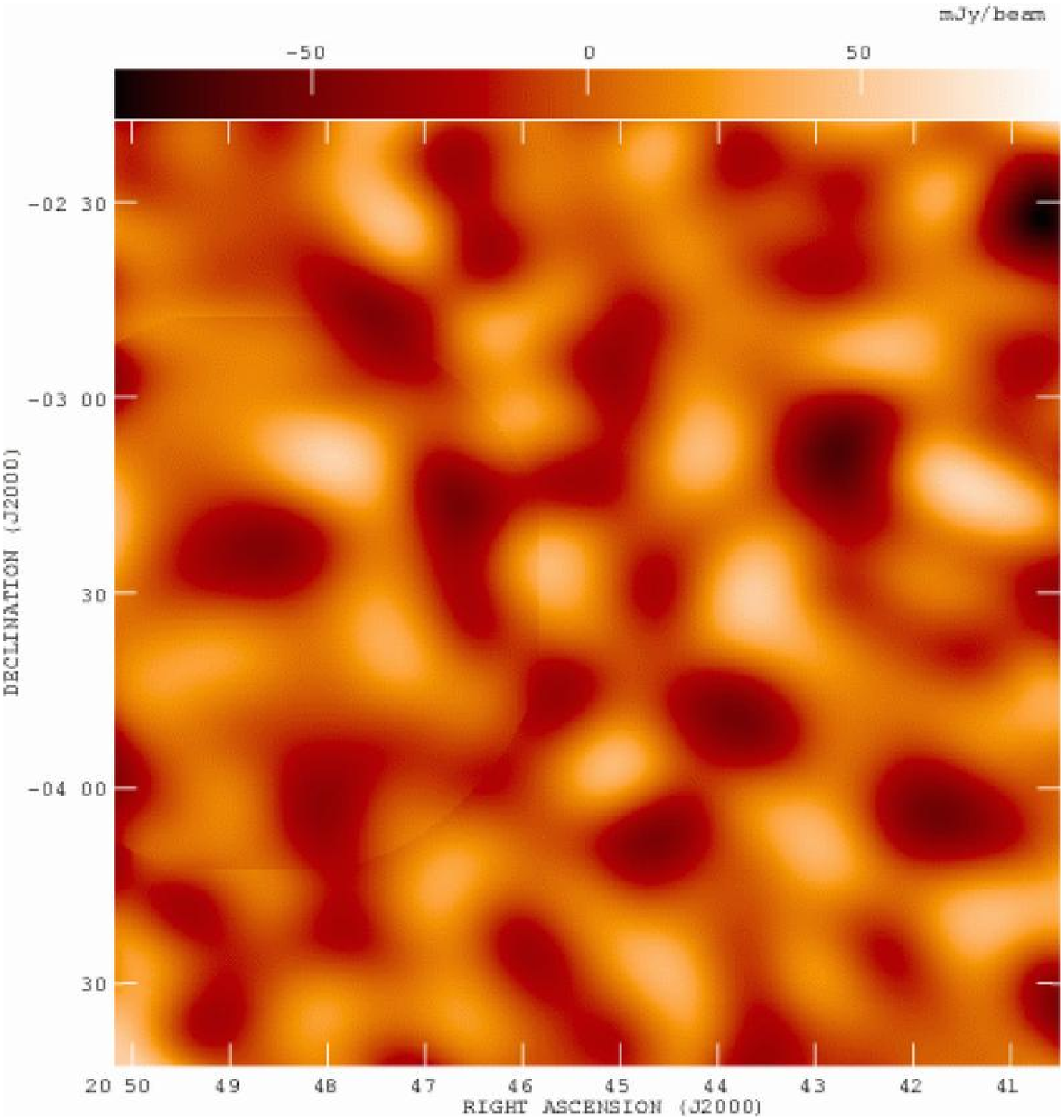}
\caption{The central region of the 42-field mosaics of differenced maps. Each map covers an area of 2.4$^{\circ}$ x 2.4$^{\circ}$. The RA scale refers to the position of the lead field. {\it Left}: VSA data {\it Right}: CBI data. {\it Top}: 02H mosaic {\it Centre}: 14H mosaic {\it Bottom}: 20H mosaic}
\label{fig:42f_mos}
\end{figure*}
\begin{figure*}
\includegraphics[width=0.41\textwidth]{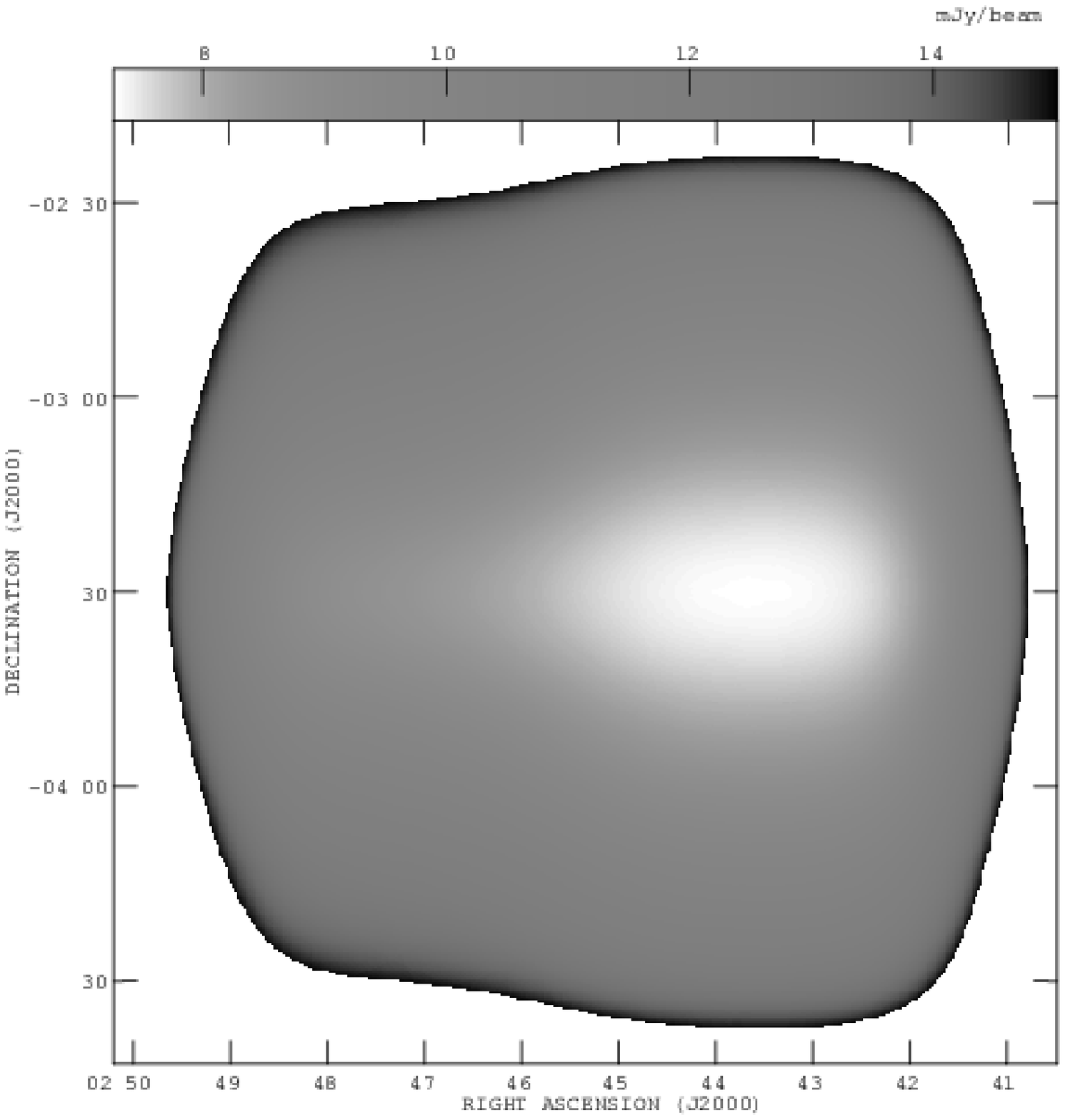}
\includegraphics[width=0.41\textwidth]{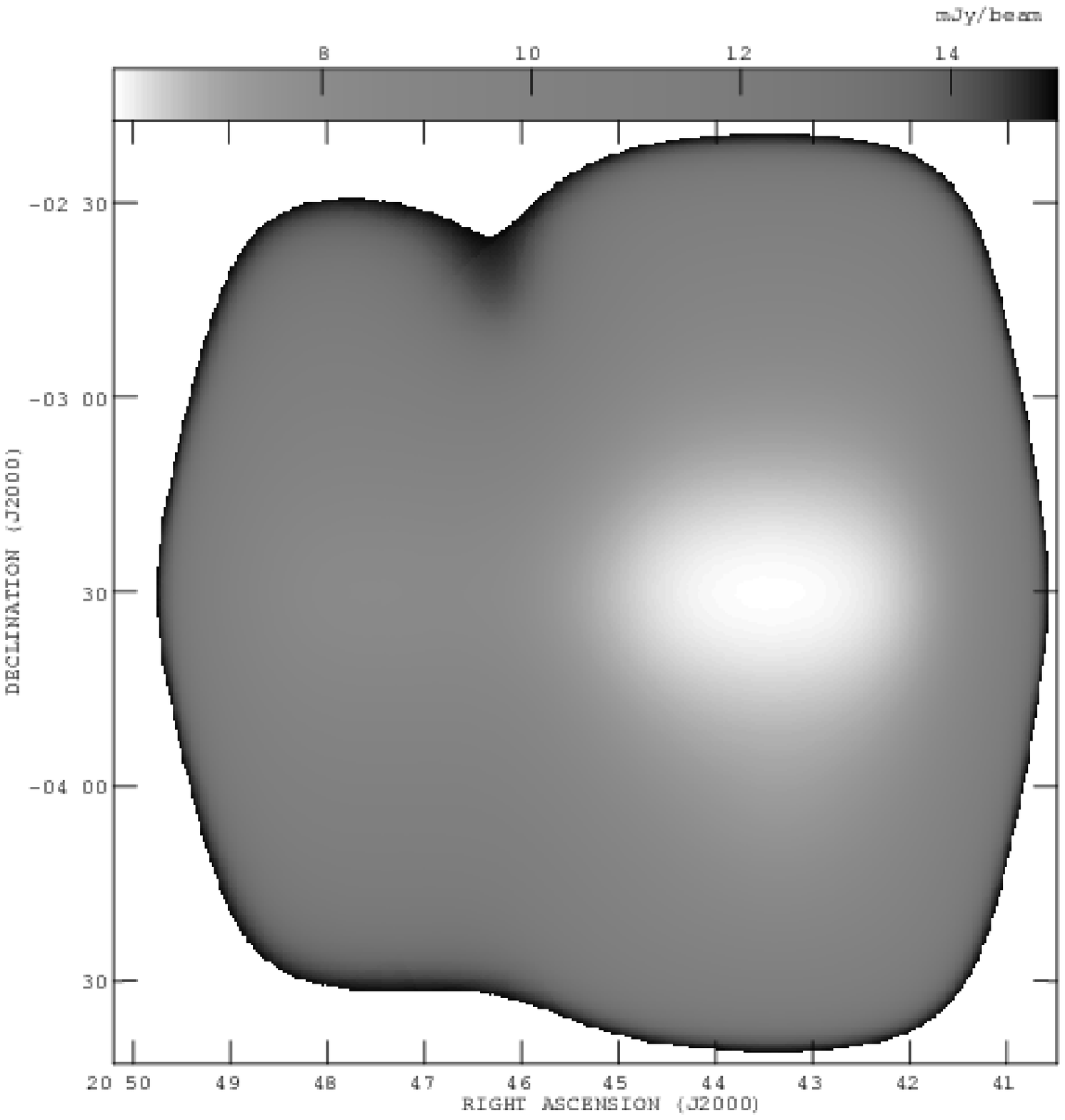}
\includegraphics[width=0.41\textwidth]{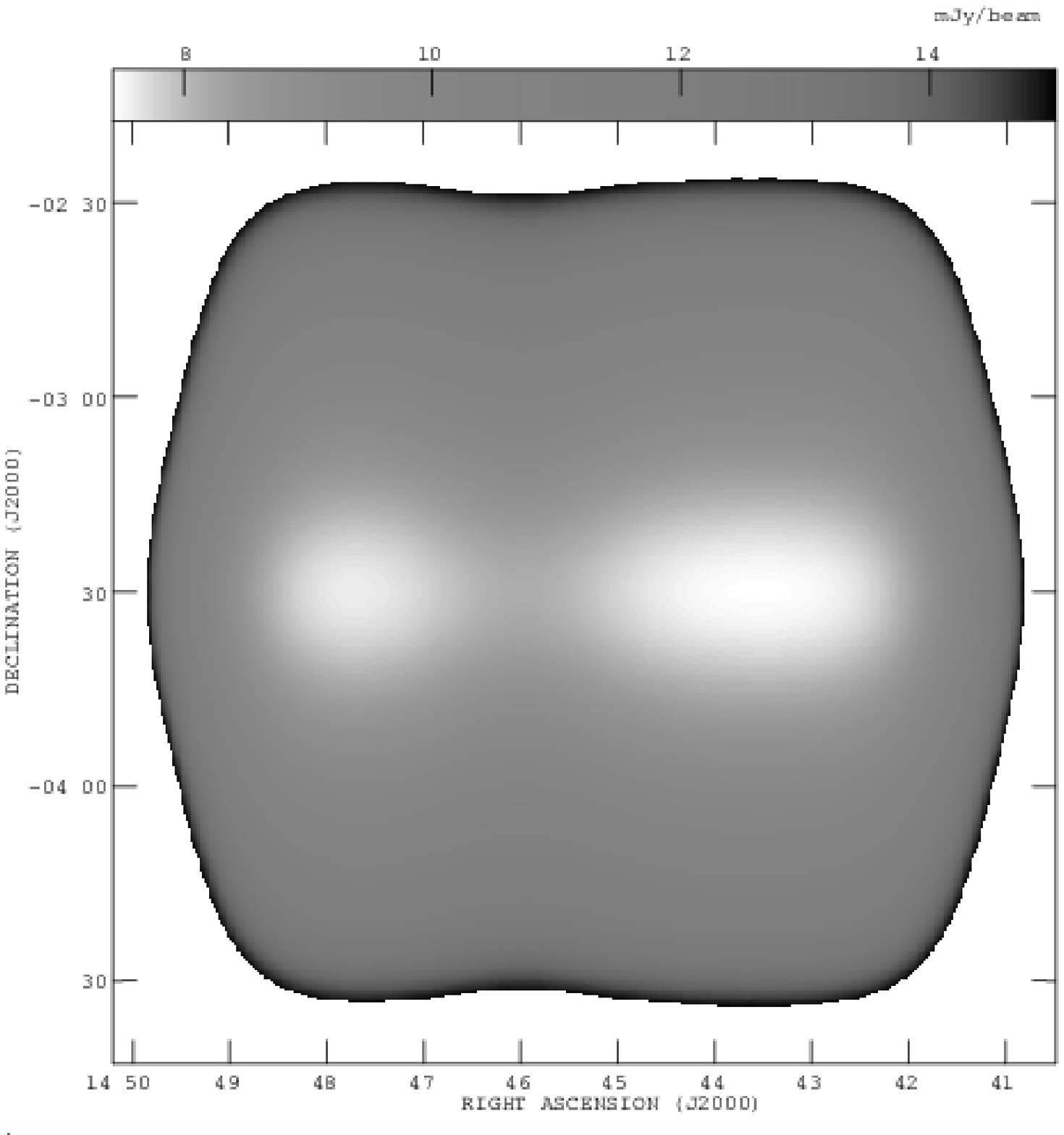}
\includegraphics[width=0.41\textwidth]{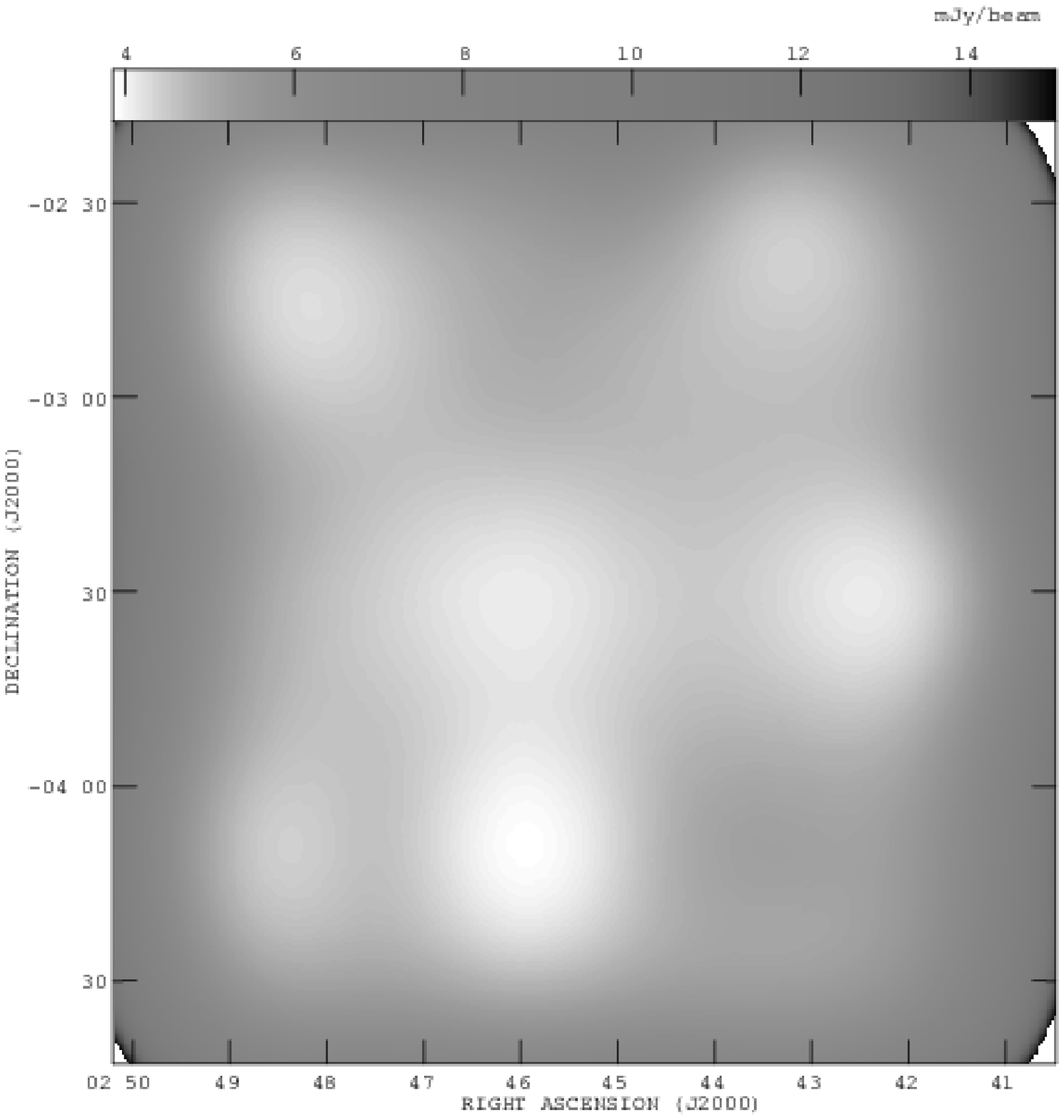}
\includegraphics[width=0.41\textwidth]{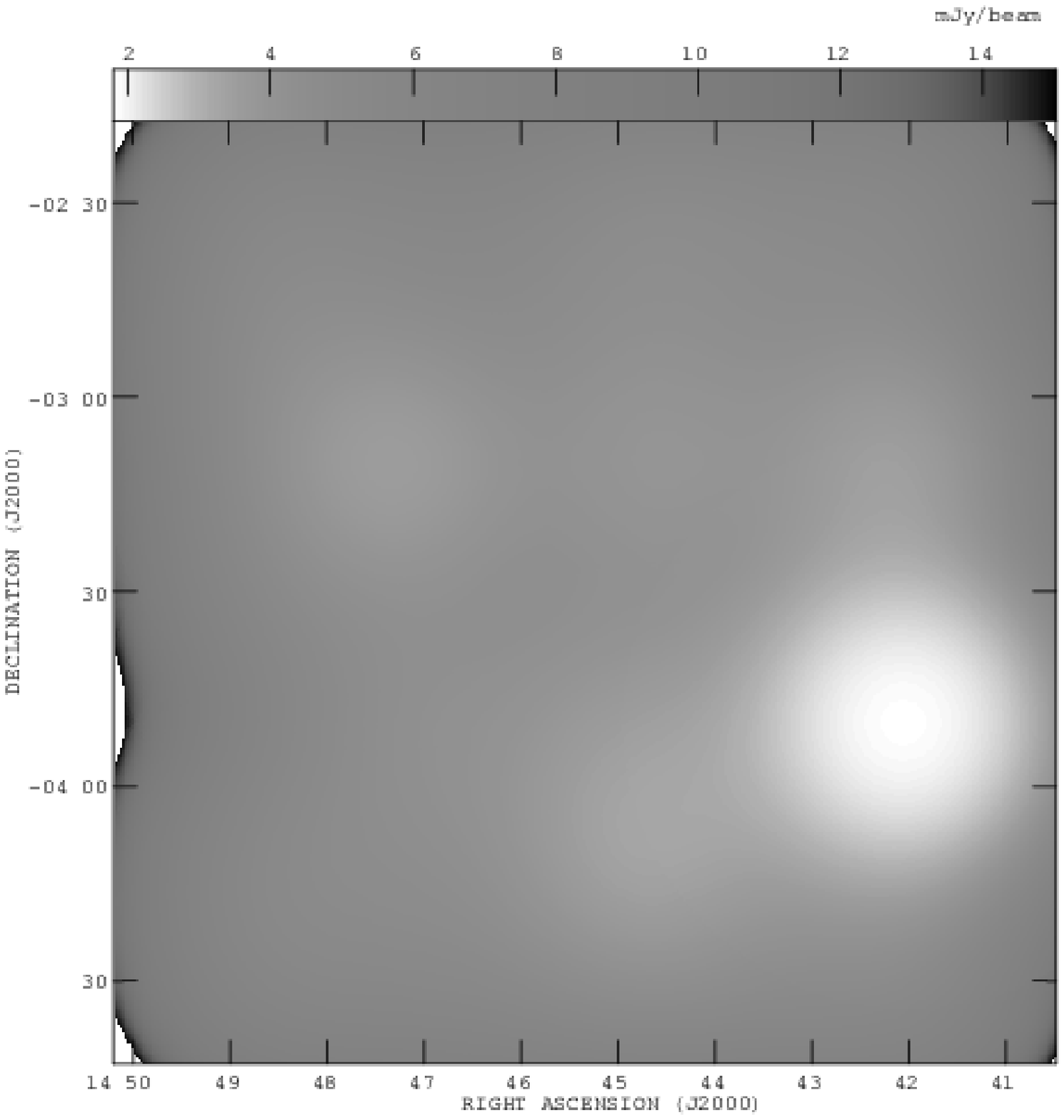}
\includegraphics[width=0.41\textwidth]{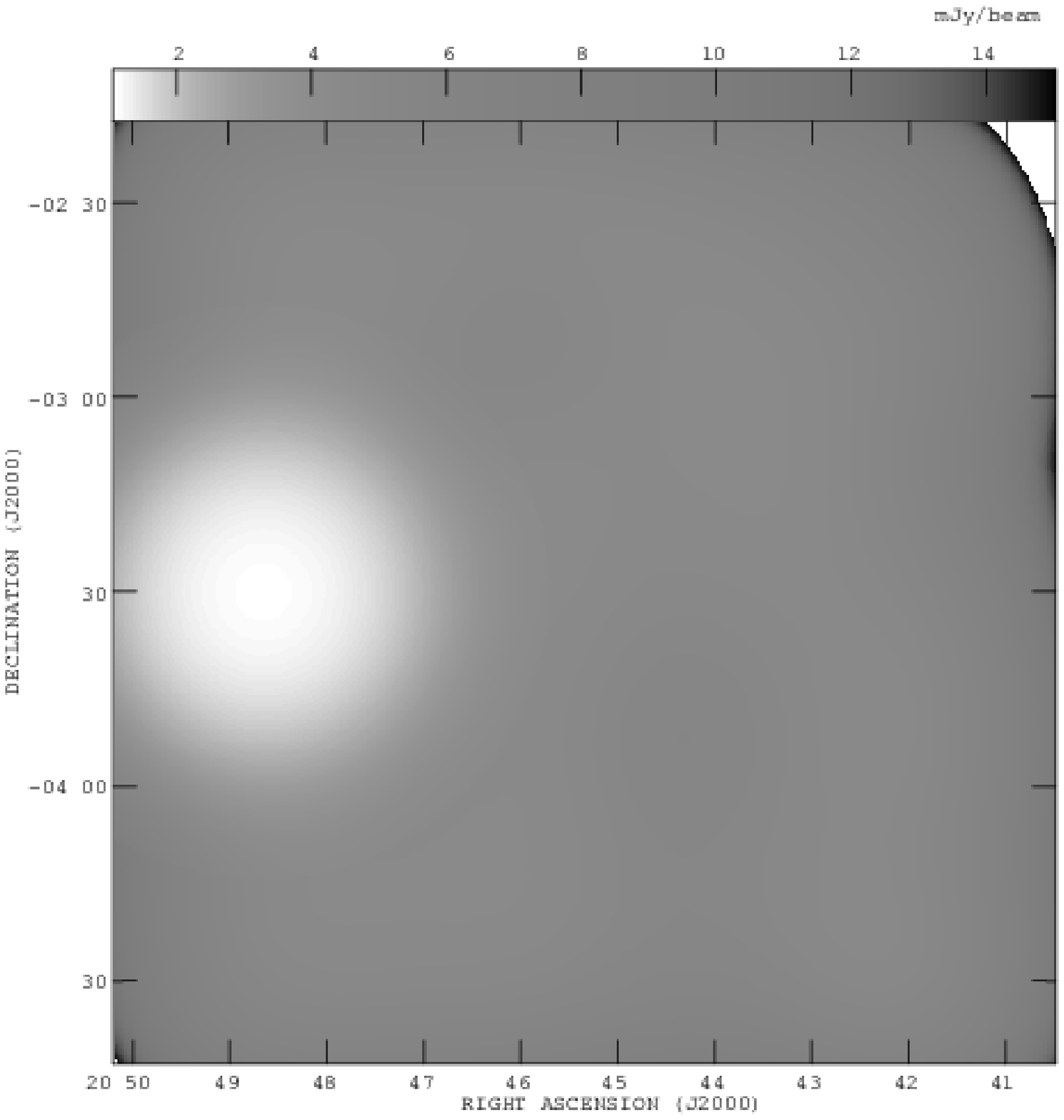}
\caption{Sensitivity maps for each mosaic shown in \ref{fig:42f_mos}. Each map covers an area of 2.4$^{\circ}$ x 2.4$^{\circ}$ and is blanked at a threshold of 15\,mJy\,beam$^{-1}$. The RA scale refers to the position of the lead field.  {\it Left}: VSA data {\it Right}: CBI data. {\it Top}: 02H mosaic {\it Centre}: 14H mosaic {\it Bottom}: 20H mosaic}
\label{fig:sens}
\end{figure*}

As a consequence of the need to difference the CBI visibilities, maps from the two telescopes may only be compared if the differencing scheme is also implemented on the VSA data. Further to the usual data reduction, the VSA data were processed is several ways to enable this. 

The VSA has a much larger primary beam than the CBI and covers the same area of sky in fewer pointings. To produce data equivalent to each CBI lead and trail field, the following procedure was implemented. The first step was to shift the VSA field centres to each of the CBI field centres. The complex visibility measured by an interferometer is defined as:

\begin{equation} \label{eq:point1}
  V(\bvec{u}) = \int d^2\bvec{x}\, {\cal A}(\bvec{x})\,I(\bvec{x})\,
   e^{-2\pi i \sbvec{u}\cdot\sbvec{x}} + {N}(\bvec{u})
\end{equation}

\noindent
where ${\cal A}(\bvec{x})$ is the primary beam, $I(\bvec{x})$ is sky intensity, $\bvec{u}=(u,v)$ is the baseline length, measured in units of the wavelength and $N(\bvec{u})$ is the instrumental noise \citep{tms...2001}. The field shift was achieved by rotating the phase of the VSA visibilities so that the direction cosines, $\bvec{x} = (\Delta x,\Delta y)$, were defined with respect to a new phase centre:

\begin{equation}
\vspace{-0.5in}\Delta x  =  \cos\delta \, \sin(\alpha - \alpha_0) \nonumber
\end{equation}
\begin{equation}
\Delta y  =  \sin\delta\,\cos\delta_0 - 
       \cos\delta\,\sin\delta_0\,\cos(\alpha - \alpha_0)\nonumber
\end{equation}

\vspace{0.17in}
\noindent
where $\alpha,\delta$ and $\alpha_0,\delta_0$ are the right ascension and declination of the VSA and CBI field centres respectively.

At this stage, before differencing can be carried out, the larger primary beam of the VSA must be taken into account. The effect of observing a limited area of sky is to convolve the sky Fourier modes with the aperture function (the Fourier transform of the primary beam).  To mimic observations with the CBI beam which has a FWHM of 45.2 arcminutes $\times$ (31\,GHz/$\nu$), the VSA visibilities were convolved with the CBI aperture function. This was modelled as a Gaussian with a cut-off at 0.45\,m (the outer radius of the antenna) and a central region with no illumination at r $<$ 0.0774\,m (corresponding to blockage by the secondary mirror). The 84 lead and trail fields for each mosaic were produced by convolving the phase rotated data with the CBI aperture function. Corresponding lead and trail fields were then differenced. 

It is important to note that matching the uv-coverage of the two interferometers is critical in making maps of the CMB. This is a consequence of sampling a random Gaussian field. Not only must the same range of angular scales be used, but they must sample the same region of uv-space. Figure \ref{fig:uv} illustrates the mismatch in uv-coverage between the VSA and CBI in the range of common uv-scales. To overcome this, the data were binned and re-weighted. As the data were obtained from a convolution of the sky fourier modes with the CBI aperture function, which has a FWHM of $67\lambda$, a cellsize of 17$\lambda$ was chosen to ensure that the aperture function was more than adequately sampled according to the Nyquist sampling theorem. Data cells with a match in both sets were assigned a weight of the geometric mean of their individual weights. Data cells adjacent to, but without a direct match, were downweighted by a factor of two. Cells without a direct or adjacent match were assigned a weight of zero.

Maps were then produced from the re-weighted visibilities using the \texttt{AIPS} package and are shown in Figure \ref{fig:42f_mos}. The CBI data have been standardised to 33\,GHz with the assumption of a blackbody spectrum. Both datasets have been corrected for the primary beam. In the case of the VSA data, the data were corrected for an effective primary beam where the effective beamsize, $\sigma_{\rm {eff}}$, is given by:

\begin{equation}
\frac{1}{\sigma_{\rm {eff}}^2} =  \frac{1}{\sigma_{\rm {C}}^2} + \frac{1}{\sigma_{\rm {V}}^2} \nonumber
\end{equation}

\noindent
where $\sigma_{\rm C}$ and $\sigma_{\rm {V}}$ are the CBI and VSA primary beam sizes respectively. The effective beam size is smaller than the CBI beam size by 5\%. This is because the sky signal measured by the VSA data has been multiplied by the CBI and VSA primary beams.

 Figure \ref{fig:sens} shows the sensitivity maps for each mosaic. The minimum noise in the VSA 02H, 14H and 20H mosaics is $7.29\,\rm mJy\,\rm beam^{-1}$, $7.44\,\rm mJy\,\rm beam^{-1}$ and $6.03\,\rm mJy\,\rm beam^{-1}$ respectively. The corresponding noise levels for the CBI mosaics are $3.88\,\rm mJy\,\rm beam^{-1}$, $1.82\,\rm mJy\,\rm beam^{-1}$ and $1.08\,\rm mJy\,\rm beam^{-1}$.  The noise levels in the CBI maps are approximately a factor of 2 lower than those of the VSA maps, with a wider gap in the region of the CBI deep fields which are clearly visible in the sensitivity maps. The blanking threshold of $15\,\rm mJy\,\rm beam^{-1}$ is for illustrative purposes only. In calculating the map correlations, a cut was applied when the power in the primary beam reached 1/e of the maximum level.

\begin{table}
\begin{center}
\caption{Correlation of the 02H and 14H mosaics. The expected  correlation is in excellent agreement with the actual correlation of the 02H and 14H mosaics.
\label{tab:cor1}}
\begin{tabular}{lcc}
\hline \hline
& Expected & 02H  \\
Map plane & $0.23 \pm 0.02$ & $0.25 \pm 0.03$ \\
uv-plane & $0.12 \pm 0.02 $ & $0.14 \pm 0.02$ \\
\hline
& Expected & 14H \\
Map plane & $0.23 \pm 0.02$ & $0.24 \pm 0.03$ \\
uv-plane & $0.12 \pm 0.02 $ & $0.13 \pm 0.02$ \\
\hline \hline
\end{tabular}
\end{center}
\end{table}

\begin{table}
\caption{Correlation of the 20H mosaic. The expected correlation in the 20H mosaic given the greater weight of data in these fields is shown in the first column, and in the second column this is modified by the inclusion of the VSA summer phase errors. The actual correlation, in column three, is consistent with the level expected given the phase calibration errors.
\label{tab:cor2}}
\begin{tabular}{lccc}
\hline \hline
& Expected & Expected  & 20H \\\vspace{0.04in}
& & (incl. phase errors) \\
Map plane & $0.27 \pm 0.02$ & $0.23 \pm 0.02$ & $0.22 \pm 0.03$\\
uv-plane & $0.14 \pm 0.02 $ & $0.11 \pm 0.02$ & $0.10 \pm 0.02$\\
\hline \hline
\end{tabular}
\end{table}

There are a number of statistics which may be used to quantify the consistency between datasets. The measure used here for both the map plane and uv-plane tests is the product-moment correlation coefficient \citep[e.g.][]{stats...1989}. 

\begin{equation}
r\,=\frac{\overline{xy}-\overline{x}\,\overline{y}}{\sigma_x \sigma_y} \nonumber
\end{equation}

In both planes a weighting scheme was used. In correlating the maps, each pixel was weighted by the power of the primary beam at the centre of the pixel. In the uv-plane, each cell was weighted by the inverse of the noise squared. As the sky is real, the underlying fourier modes a$(\bvec{u})=a^{\ast}(\bvec{-u})$. Consequently, correlations exist between visibilities that lie on opposite sides of the uv-plane. To account for this, the conjugate symmetry of the visibilities was used to reflect the data into one half of the plane and the correlation was carried out only in this region.  The real and imaginary parts of the visibilities were treated independently. 

One of the disadvantages usually cited with this measure of correlation is the difficulty of interpretation. This has been overcome by using Monte-Carlo simulations of the data to predict the expected correlation. The input model for these simulations was a standard $\Lambda$CDM model with noise levels and uv-coverage appropriate to the actual observations. The method of producing the VSA differenced data from the simulations was carried out in the same manner as for the actual observations.

The correlation of the 02H and 14H mosaics is shown in Table \ref{tab:cor1}.   There is excellent agreement between the observed and expected correlations in the 02H and 14H mosaics in both the map plane and the uv-plane. 

The expected correlation of the 20H mosaics is higher than for the 02H and 14H mosaics, which reflects the greater signal-to-noise in these observations. The actual correlation however, lies 2$\sigma$ away from the expected level - see Table \ref{tab:cor2}. In contrast to the first two mosaics, the VSA data in the 20H mosaic were largely collected during the summer period and between the hours of 8am and 6pm. In this period, the accuracy of the phase calibration is known to deteriorate and typical phase errors are some $20^{\circ}$. The source of these errors is thought to be due to a slight warping of the tilt table in the heat. All of the VSA data in the 02H and 14H mosaics were made outside of the summer daytime period, when typical phase errors are around 3-4$^{\circ}$.

The VSA phase errors are estimated from the change in phase calibration factors over the course of a day. Several calibration observations are carried out each day but there may be significant change in the phase calibration between observations. The calibrators are either point sources or resolved sources for which a model is known. The phase errors have a weak dependence on baseline length but show a high degree of repeatability. The expected correlation is, of course, revised downwards by modelling the summer phase errors in the simulations. The phase errors fully account for the reduced correlation seen in the 20H mosaics.

\begin{figure}
\includegraphics[height=8.1cm,width=5cm,angle=-90]{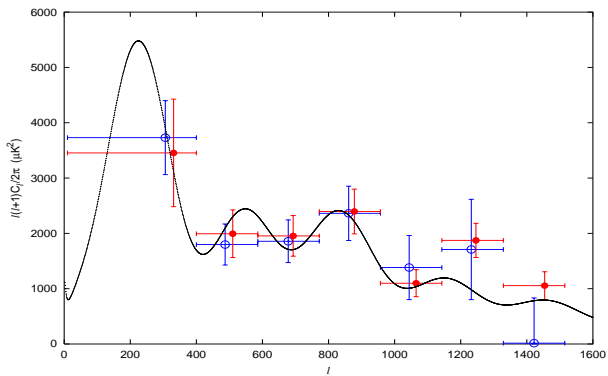}
\caption{The power spectrum of the 02H, 14H and 20H mosaics from the undifferenced VSA data (open circles) and the differenced CBI data (solid circles). The concordance model which best fits WMAP data is shown for comparison.}
\label{fig:ps}
\end{figure}

\begin{table}
\caption{The CMB band-powers ($T_0^2\ell(\ell+1)C_{\ell}/2\pi~ [\mu
{\rm K}^2]$) for the joint mosaic power spectra shown in figure \ref{fig:ps}. $\ell_{\rm eff}$ is the centroid of the bandpower window function. 
\label{tab:bands}}
\centering
\begin{tabular}{lccccc}
\hline\hline
Bin & $\ell$-range &  VSA & $\ell_{\rm eff}^{\rm VSA}$ & CBI & $\ell_{\rm eff}^{\rm CBI}$ \\
\hline
1 & $  0 - 400 $   & $3730\,\pm\,669$ &  306 & $3455\,\pm\,973  $ & 311 \\
2 & $400 - 586 $   & $1798\,\pm\,372$ &  487 & $1993\,\pm\,432  $ & 490 \\
3 & $586 - 772 $   & $1858\,\pm\,386$ &  679 & $1955\,\pm\,367  $ & 673 \\
4 & $772 - 957 $   & $2363\,\pm\,490$ &  861 & $2396\,\pm\,405 $ & 858  \\
5 & $957 - 1143 $  & $1382\,\pm\,580$ &  1044& $1099\,\pm\,244  $ & 1045 \\
6 & $1143 - 1329 $ & $1709\,\pm\,907$ &  1232& $1873\,\pm\,309 $ & 1226 \\
7 & $1329 - 1515 $ & $15\,\pm\,817$ &  1416& $1055\,\pm\, 251 $ & 1434 \\
\hline\hline
\end{tabular}
\end{table}

In principle, the power spectrum is insensitive to the phases of the visibilities, although this depends upon the level of correlation of the phase errors. The phase errors on the VSA data are found to be highly correlated. Although typical summer phase errors are some 20$^{\circ}$, the rms about the mean phase error for each baseline is only 2-3$^{\circ}$. A further consideration is the number of baselines which contribute to each uv-cell. If a large number of cells have contributions from several baselines then the phase errors will be less correlated and it is possible that this could affect the power spectrum estimate. However, simulations of VSA shallow field observations which include the effect of the crossing uv-tracks and summer phase errors show that this has a negligible effect on the power spectrum. However, as a further safeguard the VSA group discard the worst affected portion of summer data.

Figure \ref{fig:ps} shows the joint power spectrum for the 02H, 14H and 20H mosaics in the range of scales common to both instruments. The VSA power spectrum was estimated using the direct VSA observations, prior to the implementation of the differencing and reweighting scheme. There is excellent agreement in the first six bins. As the contribution of point sources increases as $\ell^2$, the discrepancy in the final bin may be due to the variability of sources, as the data were collected at different epochs. Table \ref{tab:bands} shows the binning and bandpowers for the power spectra.

\section{Window Functions}

\begin{figure}
\begin{center}
\includegraphics[height=0.5\textwidth,angle=-90,scale=1.0]{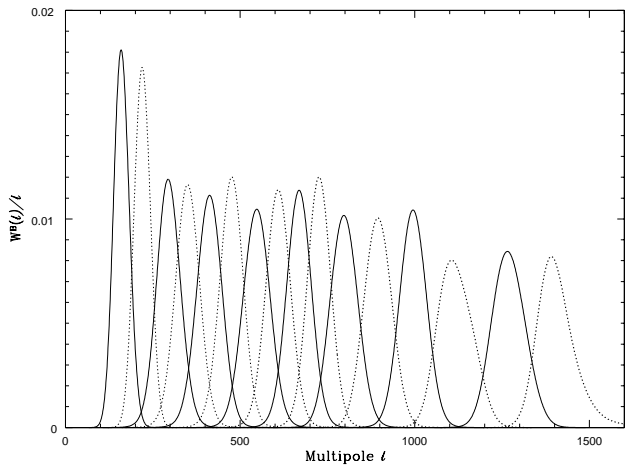}
\includegraphics[height=0.5\textwidth,angle=-90,scale=1.0]{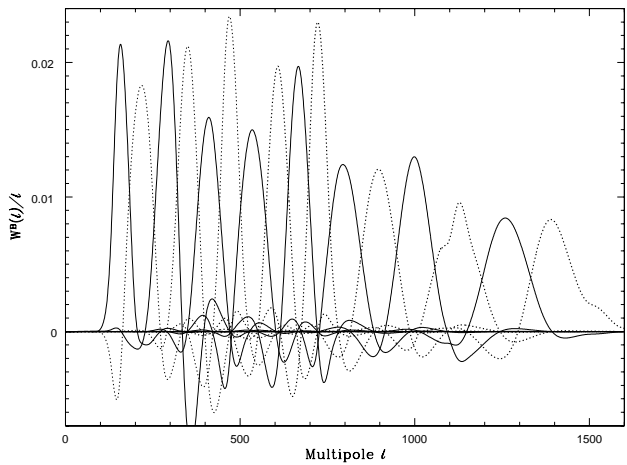}
  \caption{{\it top:} VSA variance window functions. {\it bottom:} VSA bandpower window functions. Alternate curves are solid and dotted lines for clarity. \label{fig:window_functions}}
\end{center}
\end{figure}

It is standard practice for experiments observing a limited sky area to assume a flat bandpower when estimating the CMB power spectrum \citep{2002AAS...20114004Kshort,2002MNRAS.334..569H,2003ApJ...591..575Mshort}. However, the theoretical power spectra are not flat and in estimating the cosmological parameters it is necessary to define a window function which allows a theoretical prediction of the flat bandpower $q_B$ which may be compared to the experimental values. The theoretical prediction is the expectation value of $q_B$ for a given input model: 
\begin{equation} \langle q_B \rangle = \sum_{\ell} \left( \frac{W_{\ell}^{B}}{\ell} \right)  C_{\ell} \left(a_{p} \right),  \end{equation}

\noindent
To estimate the bandpowers, the CBI group use the second order Taylor expansion of the log likelihood function around the maximum likelihood bandpowers. A quadratic approximation for the change in the bandpower, $\delta q_B$, is used iteratively, to move towards the maximum, with the second derivative of the log likelihood function replaced by its expectation value, the Fisher matrix.  The expectation value of the bandpowers is then given by

\begin{equation}
  \langle q_B \rangle  = {1\over2}\,\sum_{B'}
  \left[ F^{-1} \right]_{B B'} \,
  {\rm Tr}\left[ ( {\bf C}^{-1}\,{\bf C}^{\rm S}_{B'}\,{\bf C}^{-1}) \,
  {\bf  C}^{\rm S} \right],
\end{equation}

\noindent
where ${\bf C}$ is the covariance matrix with independent contributions from all signal sources: the CMB signal ${\bf C}^{\rm S}$, the instrumental noise ${\bf C}^{\rm N}$ and the foreground signals ${\bf C}^{\rm src}$ and ${\bf C}^{\rm res}$. ${\bf C}^{\rm S}_{B'}$ is the CMB signal from each band \citep{2003ApJ...591..575Mshort}. Since

\begin{equation}
  {\bf C}^{\rm S} \equiv \sum_{B} {\bf C}^{\rm S}_B =
  \sum_{\ell} \frac{\partial {\bf C}^{\rm S}}{\partial {\cal C}_{\ell}}
  \,{\cal C}_{\ell},
\end{equation}

\noindent
the bandpower window functions can be calculated, once the maximum likelihood bandpowers have been obtained, from

\begin{equation}\label{eq:windows}
  { W^B_{\ell} / \ell } = {1\over2}\,\sum_{B'}
  \left[ F^{-1} \right]_{B B'} \,
  {\rm Tr}\left[ ( {\bf C}^{-1}\,{\bf C}^{\rm S}_{B'}\,{\bf C}^{-1})\,
  \frac{\partial {\bf C}^{\rm S}}{\partial \cal{C}_{\ell}}\right],
\end{equation}

\noindent
where F is the Fisher matrix of the fine-binned $C_{\ell}$ estimates \citep{1999PhRvD..60j3516K}. 

The VSA group also use a maximum likelihood method to estimate the bandpowers but do not use the approximations described above. Instead the exact likelihood function is calculated as a function of each bandpower through the maximum-likelihood point, with some speed-up achieved by the use of the signal-to-noise eigenbasis \citep{2002MNRAS.334..569H}.
For parameter estimates the VSA group do not compute the bandpower window functions but use as an approximation the variance window function which is rapidly constructed from the overlap integrals of the aperture functions for pairs of visibilities and is defined as
\begin{equation}
\frac{W_{\ell}^{V}}{\ell} = \int_{0}^{2\pi} |S(\ell,\phi)|^2 \, {\rm
d}\phi ,
\end{equation}
\noindent
where $\ell=2\pi|\textbf{u}|$ and 
\begin{equation}
S(\textbf{u})=\sum_k w_k \tilde{A}_{\mbox{eff}} (\textbf{u}-\textbf{v}_k),
\end{equation}
where $w_k$ is the noise weighting of the visibilities and $\tilde{A}_{\mbox{eff}}$ is the effective aperture function \citep{2003MNRAS.341.1076Sshort}

Knox (1999) has noted that the bandpower window function is distinct from the 
variance window 
function where the signal or noise is correlated. Here we investigate the effect of 
the window function by estimating the cosmological parameters from the VSA 
dataset with each set of 
window functions. We have used the full compact and extended 
array dataset. The binning scheme and maximum likelihood bandpowers are shown in Table \ref{tab:bandpowers}. The bandpower window functions have been computed using the CBI software \citep{2003ApJ...591..575Mshort} which has been adapted for use with the VSA specifications, including the observing frequency, bandwidth and primary beam size. Figure \ref{fig:window_functions} shows the variance and window functions for the VSA. Each variance window function is normalised to unit area. The bandpower window functions are, by definition, normalised to unit area within the band limits and to zero outside of the band. The bandpower windows show negative `sidelobe' features which indicate the anti-correlations of $C_{\ell}$s in adjacent bins.

\begin{table}
\caption{The CMB band-powers (in $\mu$K$^{2}$) for the complete
VSA data set combining both compact and extended array data. 
\label{tab:bandpowers}}
\centering
\begin{tabular}{ccccc}
\hline\hline
Bin & $\ell$-range & $ \ell_{\rm eff}$ &  $T_0^2\ell(\ell+1)C_{\ell}/2\pi~ [\mu
{\rm K}^2]$ \\
\hline
1 & $100 - 190 $ & $156$ &  $3626 _{ -1150} ^{ +1616  }$ &\\
2 & $190 - 250 $ & $220$ &  $5561 _{ -1232} ^{ +1561  }$ &\\
3 & $250 - 310 $ & $281$ &  $5131 _{ -959} ^{ +1123  }$ &\\
4 & $310 - 370 $ & $333$ &  $2531 _{ -411} ^{   +438  }$ &\\
5 & $370 - 450 $ & $410$ &  $1570 _{ -219} ^{   +246  }$ &\\
6 & $450 - 500 $ & $475$ &  $1811 _{ -356} ^{   +383  }$ &\\
7 & $500 - 580 $ & $537$ &  $2212 _{ -274} ^{   +356  }$ &\\
8 & $580 - 640 $ & $611$ &  $1736 _{ -301} ^{   +356 }$ &\\
9 & $640 - 700 $ & $670$ &  $1614 _{ -301} ^{   +329  }$ &\\
10 & $700 - 750 $ & $721$ &  $1628 _{ -356} ^{   +411  }$& \\
11 & $750 - 850 $ & $794$ &  $2486 _{ -246} ^{   +301  }$& \\
12 & $850 - 950 $ & $902$ &  $1553 _{ -274} ^{   +274  }$& \\
13 & $950 - 1050$ & $987$ & $1135 _{-246} ^{  +274  }$ &\\
14 & $1050 - 1200$ & $1123$ & $677 _{ -246} ^{   +274  }$& \\
15 & $1200 - 1350$ & $1267$ & $937 _{ -329} ^{  +356  }$ &\\
16 & $1350 - 1700$ & $1440$ & $758 _{ -603} ^{  +657  }$&\\
\hline\hline
\end{tabular}
\end{table}

\begin{table}
\begin{center}
\caption{The priors assumed for the basic 
parameters. The notation $(a,b)$ for parameter $x$ denotes a top-hat prior in the
range $a<x<b$.}
\label{prior}
\begin{tabular}{lc}
\hline\hline 
Basic parameter & Prior \\
\hline
$\omega_{\rm b}$ & $(0.005,0.100)$ \\
$\omega_{\rm dm}$ & $(0.01,0.99)$ \\
$\theta$ &  $(0.5,10.0)$ \\
$\tau$ & $(0.01,0.50)$\\
$n_{\rm s}$ &  $(0.5,1.5)$ \\
$\rm ln(10^{10} A_{\rm s}$) & $(2.7,4.6)$ \\
\hline\hline
\end{tabular}
\end{center}
\end{table}

\subsection{Parameter Estimation}

We consider the set of cosmological models described by the following six free parameters: the physical baryon density
$\omega_{\rm b}\,=\,\Omega_{\rm b}h^2$; the physical dark matter density
$\omega_{\rm dm}\,=\,\Omega_{\rm dm}h^2$; the ratio of the sound horizon to the 
angular diameter distance $\theta$; the optical depth to the surface of last scattering $\tau$; the amplitude of scalar modes $A_s$ and the spectral index 
of scalar modes $n_s$. We assume spatial flatness, setting the curvature 
density $\Omega_{\rm k}\,=\,1-\Omega_{\rm tot}\,=\,0$. We also set  
$w\,=\,-1$ describing the equation of state of dark energy ($p = w\rho$) and do not consider tensor modes or massive neutrinos in this analysis.

The parameter estimation was performed using the October 2004 version of the \texttt{CosmoMC} software package \citep{2002PhRvD..66j3511L}. This uses a new parameterisation with $\theta$ instead of $\rm H_{\rm 0}$ as a basic parameter. As $\theta$ is less correlated with other parameters than $\rm H_{\rm 0}$, this has the advantage of allowing the Markov chains to converge more quickly. 

In addition to the priors on the basic parameters, listed in Table~\ref{prior}, we also impose priors on some of the derived parameters. Specifically, we use top-hat priors on the age of the Universe lying between 10 and 20 Gyr, and of $\rm H_{\rm 0}$ lying between 40\,km\,s$^{-1}$\,Mpc$^{-1}$ and 100\,km\,s$^{-1}$\,Mpc$^{-1}$.

\begin{figure*}
\includegraphics[width=13cm,angle=0]{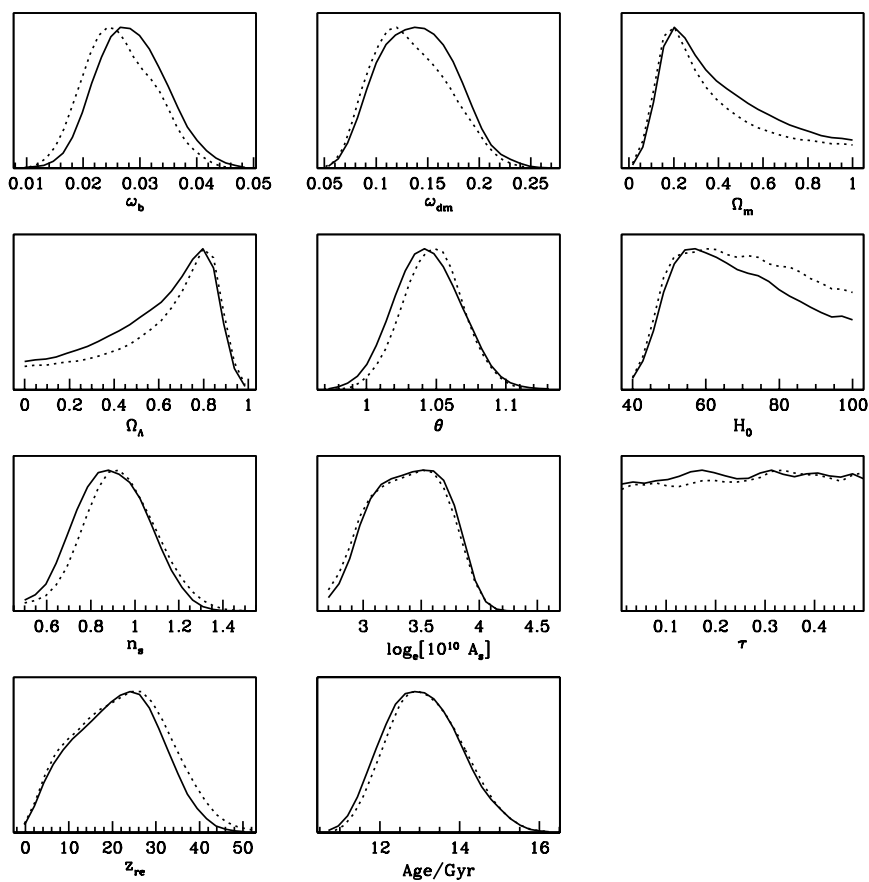}
\caption{The one-dimensional marginalised probability distributions for
cosmological parameters estimated using variance windows (solid lines) and 
bandpower windows (dotted lines).}
\label{fig:mar}
\end{figure*}

\begin{table}
\begin{center}
\caption{Parameter estimates and 68\% confidence intervals from the marginalised distributions. When the marginalised 
posterior for a parameter does not contain a peak, 95\% confidence limits 
are given.
\label{tab:res1}}
\begin{tabular}{lcc}
\hline\hline
&  Variance windows & Bandpower windows  \\
\hline
$\omega_{\rm b}$ & $0.029\pm ^{ 0.006 }_{ 0.006}$ & $0.027\pm ^{ 0.006 }_{ 0.006}$  \\

$\omega_{\rm dm}$ &   $0.14\pm ^{ 0.04 }_{ 0.04}$ & $0.13\pm ^{ 0.04}_{ 0.04}$ \\

$\Omega_{\rm m}$ &   $0.42\pm ^{ 0.09 }_{ 0.17}$ & $0.39\pm ^{ 0.07 }_{ 0.17}$  \\

$ \Omega_{\Lambda}$ & $0.58\pm ^{ 0.17 }_{ 0.09}$ &  $0.61\pm ^{ 0.17 }_{ 0.07}$   \\

$\theta$ & $1.04\pm ^{0.02}_{0.02}$ & $1.05\pm ^{0.02}_{0.02}$\\

$\rm h$ & $ 0.69\pm ^{  0.10 }_{  0.10}$ &  $ 0.71\pm ^{  0.09 }_{  0.10}$  \\

$ n_{\rm s}$ & $0.90\pm ^{ 0.07 }_{ 0.07}$ &  $0.93 \pm ^{ 0.07 }_{ 0.08}$  \\

$ \rm ln 10^{10} A_{\rm s}$ &   $3.4\pm ^{ 0.2 }_{ 0.2 }$ & $3.4 \pm ^{0.2}_{0.2}$  \\

$\tau$ & $(0.04, 0.47)$  & $ (0.04, 0.48)$  \\

$z_{\rm re}$ &  $21\pm ^{10}_{10}$  & $22\pm ^{11}_{11}$ \\

$\mbox{Age}$ &   $13.1\pm ^{ 0.9 }_{ 0.9}$ &  $13.2\pm ^{0.9}_{0.9}$  \\

\hline\hline
\end{tabular}
\end{center}
\end{table}

The \texttt{CosmoMC} software was run on a 24-node linux cluster. The chains were run until the largest eigenvalue returned by the Gelman-Rubin convergence test reached 0.08. After burn-in a total of more than 100,000 samples were collected. As successive samples in a Markov chain are, by nature, correlated, the samples were thinned by a factor of 25 resulting in approximately 4000 independent samples. These samples were then used to calculate the marginalised distributions and parameter estimates (see table \ref{tab:res1}).

Figure \ref{fig:mar} shows the marginalised distributions for each parameter. It is clear that no significant bias has been introduced as a result of using variance windows to approximate bandpower windows. The largest discrepancy is seen in the parameter estimate for $\omega_{b}$ where the bandpower windows reduce the estimate by one-third of the 1-sigma error. For all other parameters the estimates are consistent to a much smaller fraction of the error. The width of each distribution is also shown to be unchanged by any significant amount. Furthermore, the correlations of the parameter estimates are also consistent using both methods. 

These results are perhaps to be expected given the sparse nature of the covariance matrix. For the VSA extended array data, approximately 5\% of the elements of the matrix are non-zero and in the limit of a diagonal covariance matrix the bandpower and variance windows are equivalent. However, the validity of the approximation also depends on the signal-to-noise level and on the binning used. The level of correlations between the bins and the gradient of the power spectrum across the bin also has an impact on the bandpower predicted. For example, using a standard $\Lambda$CDM power spectrum and the binning used above, we can compare the bandpowers predicted from each window function. In the majority of bins, the agreement is within 5\% but there is a 13\% difference in bin 3 where the gradient of the power spectrum is steep and there is a strong anti-correlation with bin 4. The effect of doubling the size of the bins increases the discrepancy between the bandpower predictions to a maximum of 28\%. Differences of this size would be likely to bias the parameter estimates. However, the level of agreement between the bandpower predictions is not a simple function of bin size and will fluctuate depending on the exact binning used. 

The VSA is currently undergoing an upgrade to increase the sensitivity of the instrument and to enable a measurement of the power spectrum up to $\ell\,=\,2500$.  The upgrade will involve the fitting of larger mirrors with a diameter of 0.55\,m. In this case, the fraction of non-zero elements in the covariance matrix will increase slightly to 7.5\%. The suitability of the approximation also depends upon the signal-to-noise achieved. The upgraded VSA will have increased sensitivity, but taken in conjunction with the reduced signal, then the overall signal-to-noise level will remain about the same. Given the significant difference between the expected bandpower values which may arise and to ensure that this does not bias future VSA parameter estimates, analysis of the upgraded VSA data will be carried out using bandpower window functions.

\section{Conclusions}

In this paper we have presented three sets of coincident CMB observations observed at different epochs by the VSA and CBI telescopes. We have chosen to analyse the full datasets, rather than focusing on the power spectra, in order to investigate any possible systematic effects which may not otherwise be revealed. The correlation of the datasets from each group was found to be as expected for the 02H and 14H mosaics. In the third mosaic the data disagreed with the Monte Carlo simulations at a level of 2$\sigma$. However, this is consistent with the phase calibration errors expected from VSA data during the summer months. It has been established that this does not affect the power spectrum estimation due to the correlated nature of the phase errors. The results of this analysis reaffirm that both groups have correctly characterised the noise properties and systematics of the telescopes as well as other, potential data contaminants.

We have investigated the use of variance windows as an approximation to bandpower windows, for the VSA, and found that for the data obtained so far, this is a valid approximation. We note that alternative binning schemes may reduce the suitability of this method and plan to use bandpower window functions for parameter estimation  from super-extended VSA data.

\section*{ACKNOWLEDGEMENTS} 

We thank Sarah Smith for useful discussions. We thank the staff of the Mullard Radio Astronomy Observatory, the Jodrell Bank Observatory and the Teide Observatory for invaluable assistance in the commissioning and operation of the VSA. The VSA is supported by PPARC and the IAC. N.Rajguru, A.Scaife, K.Lancaster   and R. S. Savage acknowledge the support of PPARC studentships. A. Slozar acknowledges the support of St. Johns College, Cambridge. G. Rocha acknowledges a Leverhulme Fellowship at the University of Cambridge.

\label{lastpage}
\bibliographystyle{mn2e}

\bsp

\end{document}